\documentclass[sigconf]{acmart}
\AtBeginDocument{%
  }

\usepackage[utf8]{inputenc} % allow utf-8 input
\usepackage[T1]{fontenc}    % use 8-bit T1 fonts
\usepackage{url}            % simple URL typesetting
\usepackage{booktabs}       % professional-quality tables
\usepackage{amsfonts}       % blackboard math symbols
\usepackage{nicefrac}       % compact symbols for 1/2, etc.
\usepackage{microtype}      % microtypography
\usepackage{xcolor}         % colors
\usepackage{comment}
\usepackage{amsmath}
\usepackage{amsfonts}
\usepackage{amssymb}
\usepackage{bbding}
\usepackage{enumitem}
\usepackage{booktabs}
\usepackage{multirow}
\usepackage{soul}  % 替代ulem
\usepackage{threeparttable}
\usepackage{array}          % for table
\usepackage{colortbl}       % for table
\usepackage{makecell}       % for table
\usepackage{tabularx}       % for table
\usepackage{booktabs}       % for tables
\usepackage{bigstrut}       % for bigstrut
\usepackage{graphicx}
\usepackage{subcaption}     % for subfigures
\usepackage{cleveref}

\newcommand{\ie}{\emph{i.e., }}
\newcommand{\eg}{\emph{e.g., }}

\newcommand{\etc}{\emph{etc}}

\newcommand{\cf}{\emph{cf. }}

\copyrightyear{2026}
\acmYear{2026}
\setcopyright{cc}
\setcctype{by}
\acmConference[WWW '26]{Proceedings of the ACM Web Conference 2026}{April 13--17, 2026}{Dubai, United Arab Emirates}
\acmBooktitle{Proceedings of the ACM Web Conference 2026 (WWW '26), April 13--17, 2026, Dubai, United Arab Emirates}
\acmPrice{}
\acmDOI{10.1145/3774904.3792387}
\acmISBN{979-8-4007-2307-0/2026/04}

\acmSubmissionID{rfp1942}

\begin{document}

%%
%% The "title" command has an optional parameter,
%% allowing the author to define a "short title" to be used in page headers.
\title{Field Matters: A Lightweight LLM-enhanced Method for CTR Prediction}

%%
%% The "author" command and its associated commands are used to define
%% the authors and their affiliations.
%% Of note is the shared affiliation of the first two authors, and the
%% "authornote" and "authornotemark" commands
%% used to denote shared contribution to the research.
\author{Yu Cui}
\authornotemark[2]
\authornotemark[3]
\orcid{0009-0001-6203-3022}
\affiliation{%
  \institution{Zhejiang University}
  \city{Hangzhou}
  \country{China}}
\email{cuiyu23@zju.edu.cn}

\author{Feng Liu}
\orcid{0009-0004-9265-9431}
\affiliation{%
  \institution{OPPO Research Institute}
  \city{Shenzhen}
  \country{China}}
\email{liufeng4hit@gmail.com}

\author{Jiawei Chen}
\authornote{Corresponding author.}
\authornote{State Key Laboratory of Blockchain and Data Security, Zhejiang University.}
\authornote{College of Computer Science and Technology, Zhejiang University.}
\authornote{Hangzhou High-Tech Zone (Binjiang) Institute of Blockchain and Data Security.}
\orcid{0000-0002-4752-2629}
\affiliation{%
  \institution{Zhejiang University}
  \city{Hangzhou}
  \country{China}}
\email{sleepyhunt@zju.edu.cn}

\author{Xingyu Lou}
\orcid{0009-0003-3180-0668}
\affiliation{%
  \institution{OPPO Research Institute}
  \city{Shenzhen}
  \country{China}}
\email{louxingyu@oppo.com}

\author{Changwang Zhang}
\orcid{0009-0004-4193-7833}
\affiliation{%
  \institution{OPPO Research Institute}
  \city{Shenzhen}
  \country{China}}
\email{changwangzhang@foxmail.com}

\author{Jun Wang}
\orcid{0000-0002-0481-5341}
\affiliation{%
  \institution{OPPO Research Institute}
  \city{Shenzhen}
  \country{China}}
\email{junwang.lu@gmail.com}

\author{Yuegang	Sun}
\orcid{0009-0009-2701-4641}
\affiliation{%
  \institution{Intelligence Indeed}
  \city{Hangzhou}
  \country{China}}
\email{bulutuo@i-i.ai}

\author{Xiaohu Yang}
\authornotemark[2]
\orcid{0000-0003-4111-4189}
\affiliation{%
  \institution{Zhejiang University}
  \city{Hangzhou}
  \country{China}}
\email{yangxh@zju.edu.cn}

\author{Can	Wang}
\authornotemark[2]
\authornotemark[4]
\orcid{0000-0002-5890-4307}
\affiliation{%
  \institution{Zhejiang University}
  \city{Hangzhou}
  \country{China}}
\email{wcan@zju.edu.cn}

%%
%% By default, the full list of authors will be used in the page
%% headers. Often, this list is too long, and will overlap
%% other information printed in the page headers. This command allows
%% the author to define a more concise list
%% of authors' names for this purpose.
\renewcommand{\shortauthors}{Cui et al.}

%%
%% The abstract is a short summary of the work to be presented in the
%% article.
\begin{abstract}
Click-through rate (CTR) prediction is a fundamental task in modern recommender systems. In recent years, the integration of large language models (LLMs) has been shown to effectively enhance the performance of traditional CTR methods. However, existing LLM-enhanced methods often require extensive processing of detailed textual descriptions for large-scale instances or user/item entities, leading to substantial computational overhead. To address this challenge, this work introduces LLaCTR, a novel and lightweight LLM-enhanced CTR method that employs a field-level enhancement paradigm. Specifically, LLaCTR first utilizes LLMs to distill crucial and lightweight semantic knowledge from small-scale feature fields through self-supervised field-feature fine-tuning. Subsequently, it leverages this field-level semantic knowledge to enhance both feature representation and feature interactions. In our experiments, we integrate LLaCTR with six representative CTR models across four datasets, demonstrating its superior performance in terms of both effectiveness and efficiency compared to existing LLM-enhanced methods. Our code is available at {\url{https://github.com/istarryn/LLaCTR}}.
\end{abstract}

%%
%% The code below is generated by the tool at http://dl.acm.org/ccs.cfm.
%% Please copy and paste the code instead of the example below.
%%


%%
%% Keywords. The author(s) should pick words that accurately describe
%% the work being presented. Separate the keywords with commas.
%% A "teaser" image appears between the author and affiliation
%% information and the body of the document, and typically spans the
%% page.
% \begin{teaserfigure}
%   \includegraphics[width=\textwidth]{sampleteaser}
%   \caption{Seattle Mariners at Spring Training, 2010.}
%   \Description{Enjoying the baseball game from the third-base
%   seats. Ichiro Suzuki preparing to bat.}
%   \label{fig:teaser}
% \end{teaserfigure}

% \received{20 February 2007}
% \received[revised]{12 March 2009}
% \received[accepted]{5 June 2009}

%%
%% This command processes the author and affiliation and title
%% information and builds the first part of the formatted document.
\maketitle

\section{Introduction}
Click-through rate (CTR) prediction \cite{yang2022click} is a core task in modern recommender systems (RS) \cite{chen2023bias,zhang2025advancing}, with the objective of estimating the likelihood that a user will click on a specific item by modeling complex feature interactions. With the advent of Large Language Models (LLMs) and their remarkable abilities in content comprehension and semantic reasoning \cite{achiam2023gpt, dubey2024llama}, there has been a surge of interest in leveraging LLMs for CTR prediction. Recent studies mainly explored two paradigms: 1)  LLMs as CTR predictors \cite{geng2022recommendation, bao2023tallrec, geng2024breaking}, where LLMs are either prompted or fine-tuned to directly perform CTR prediction; 2) LLM-enhanced CTR models \cite{li2023ctrl, xi2024towards, sun2024large, qiu2024ease}, where LLMs are leveraged to augment traditional CTR models by injecting semantic knowledge. 

\begin{figure}[t]
    \centering
    \begin{subfigure}[b]{0.46\textwidth}
        \includegraphics[width=\textwidth]{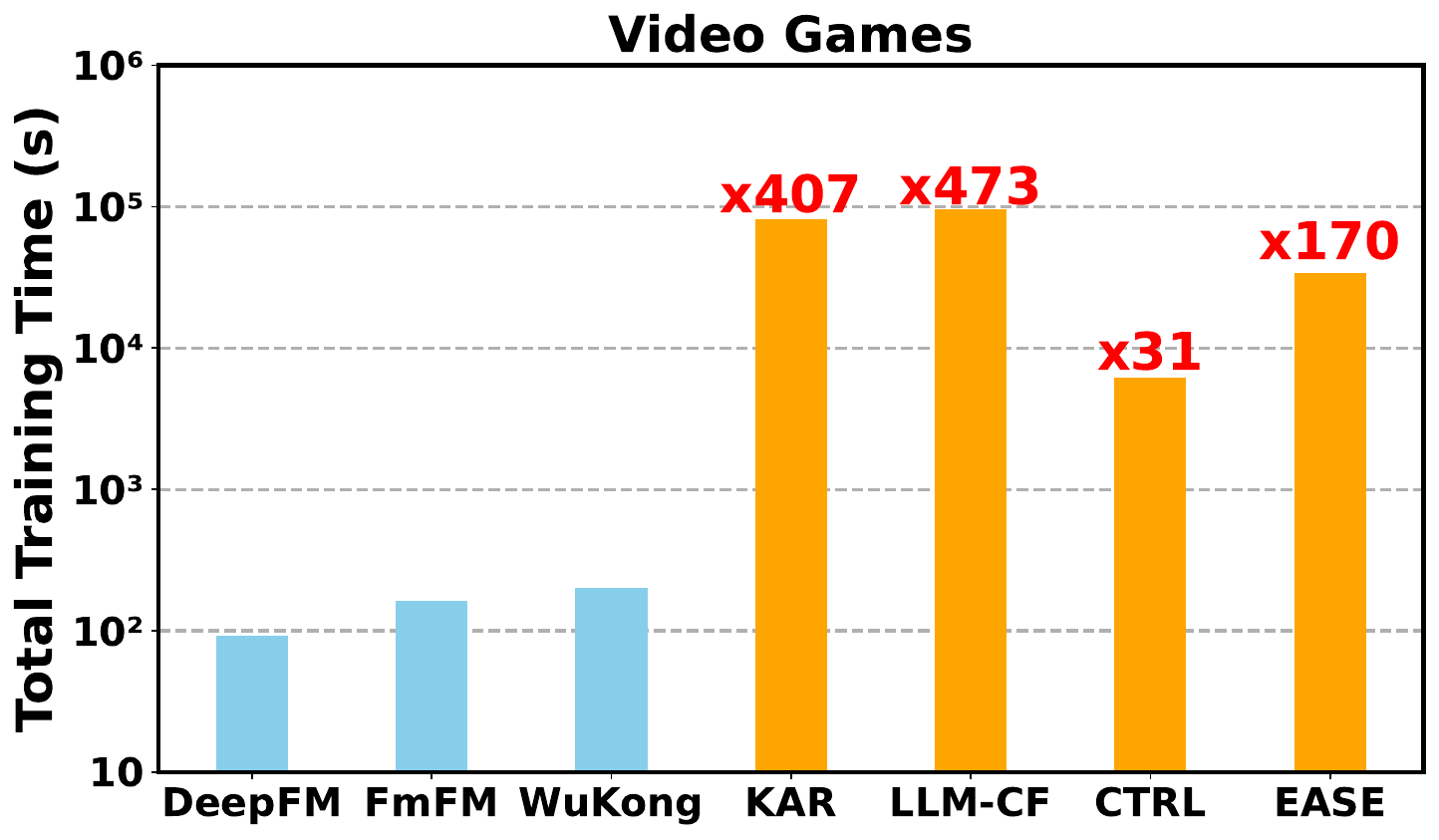}
        % \caption{Video Games}
    \end{subfigure}
    \begin{subfigure}[b]{0.46\textwidth}
        \includegraphics[width=\textwidth]{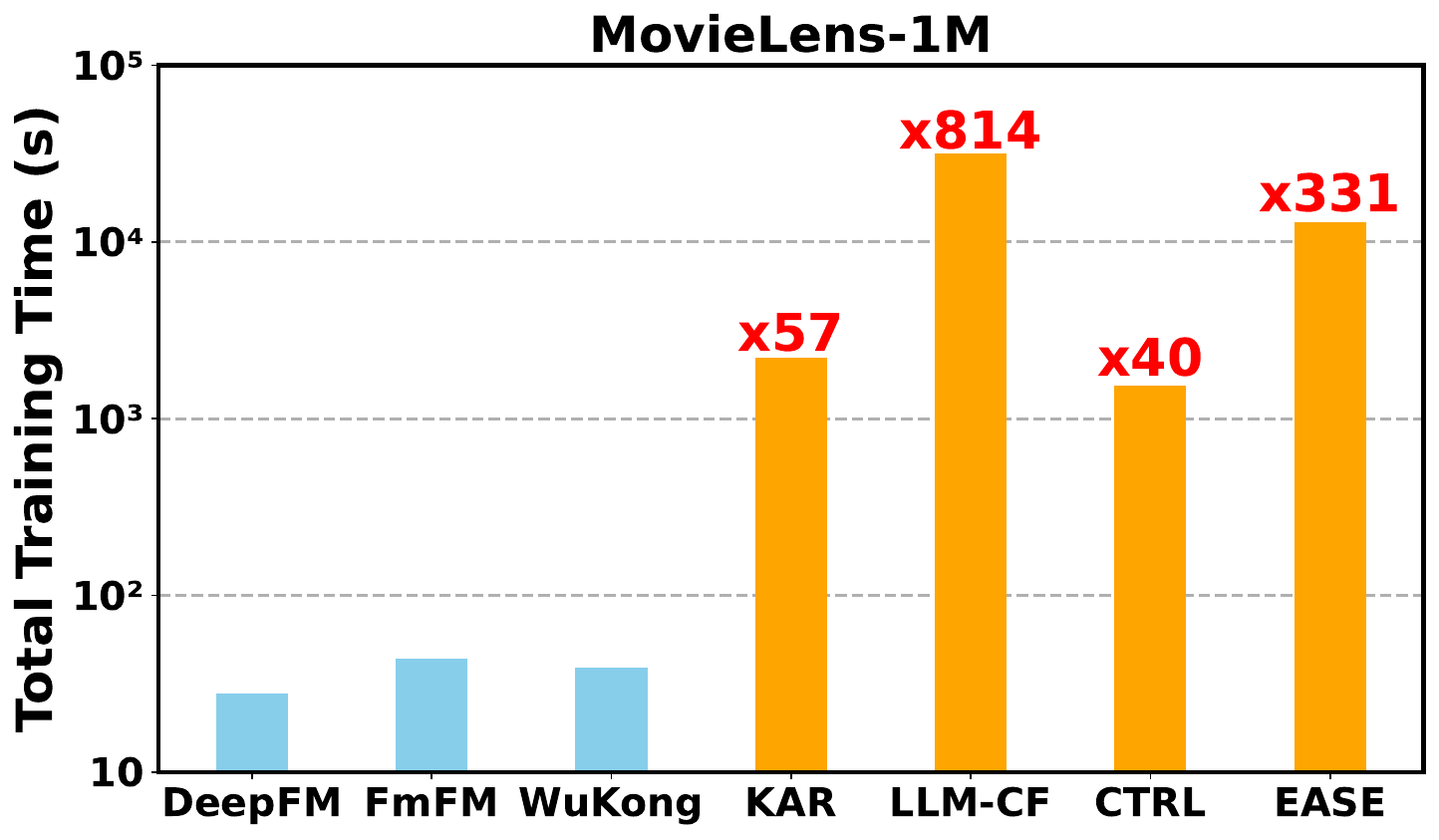}
        % \caption{MovieLens-1M}
    \end{subfigure}
    \vspace{-0.3cm}
    \caption{Empirical efficiency study on representative conventional CTR models and LLM-enhanced CTR models. The multiple increase in time cost is reported based on WuKong.} 
    \vspace{-0.5cm}
    \label{fig:motivation_empirical}
\end{figure}

Despite their encouraging performance, these approaches encounter significant practical limitations, particularly regarding computational efficiency and economic feasibility. The high computational demands of LLMs make their direct deployment for online CTR prediction largely impractical, as they introduce considerable inference latency that violates real-time serving requirements.  Although the second paradigm alleviates inference latency by retaining conventional CTR models for online serving, it still suffers prohibitive training costs. Specifically, these methods typically operate at the instance level \cite{li2023ctrl,sun2024large, lin2024clickprompt} or the user/item level \cite{xi2024towards, wang2024cela, qiu2024ease}, requiring LLMs to process detailed textual descriptions for large-scale data instances or user/item entities. Empirically, we find that existing LLM-enhanced CTR methods (\eg KAR\cite{xi2024towards}, LLM-CF\cite{sun2024large}, CTRL\cite{li2023ctrl} and EASE \cite{qiu2024ease}) require over 290 times (on average) more computation time than baseline models on typical Amazon Video Games and MovieLens-1M datasets with millions of interactions (\cf Figure~\ref{fig:motivation_empirical}). 
% This computational burden becomes even more pronounced when scaling to industrial settings involving billions of instances. 
These challenges naturally motivate a critical research question: \textbf{\emph{How can we leverage LLMs to enhance CTR models in a more efficient and cost-effective manner?}}

\begin{figure*}[t]
    \centering 
    \includegraphics[width=0.98\textwidth]{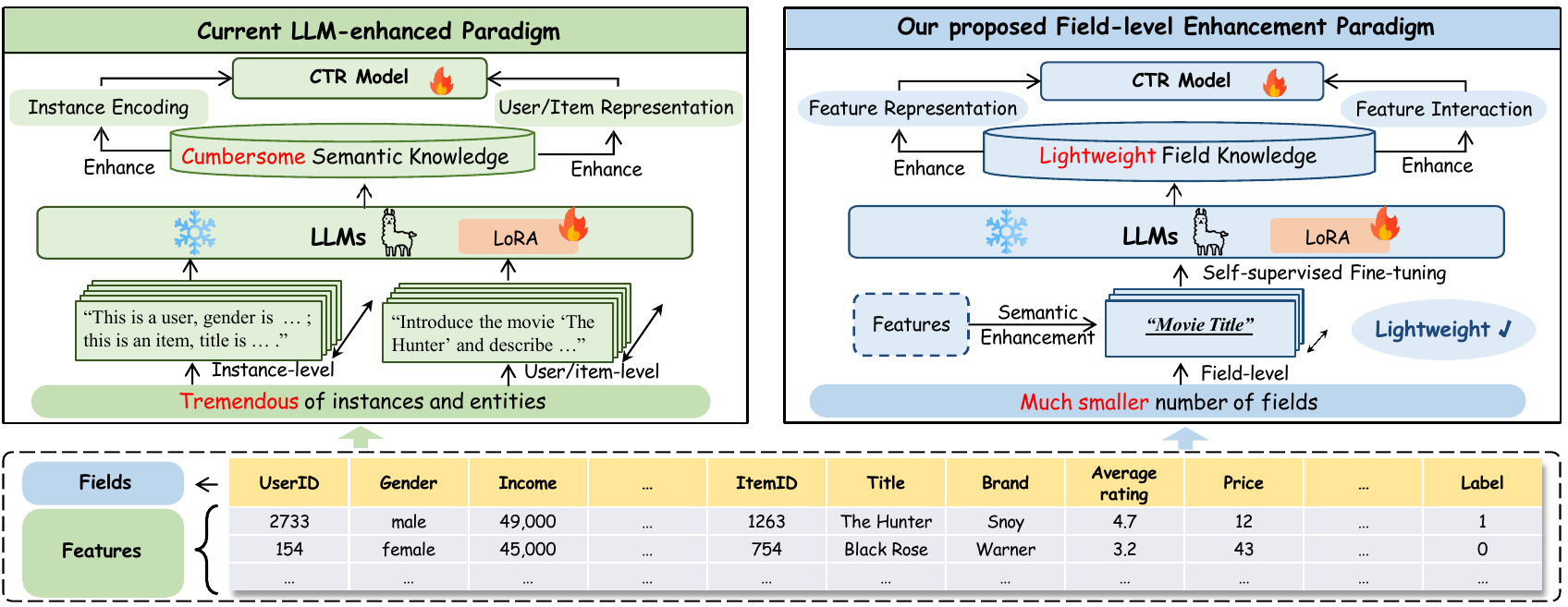}
    \vspace{-0.3cm}
    \caption{Current LLM-enhanced CTR paradigm versus our field-level enhancement paradigm.} 
    \label{fig:field}
    \vspace{-0.3cm}
\end{figure*}

To address this challenge, we explore a novel field-level enhancement paradigm that utilizes LLMs to extract semantic knowledge from feature fields to enhance CTR models. As illustrated in Figure~\ref{fig:field}, fields denote categories of features in CTR prediction. This paradigm offers two principal advantages: 1) \textbf{Computational Efficiency}: The number of fields is orders of magnitude lower than that of instances or entities (\eg hundreds vs. millions in the real RS), yielding a drastic reduction in LLM computational overhead. 2) \textbf{Crucial Semantic Information}: Conventional CTR methods primarily utilize field IDs as categorical indicators, often neglecting the rich semantic information inherent in field descriptions.  In fact, field semantics can significantly enhance both feature representation and feature interaction modeling --- the two core components contributing the success of CTR prediction. For example, recognizing that the feature “4.7” belongs to the field “average rating” provides insight into item quality; understanding the relationship between fields such as “user income” and “item price” can inform the importance of feature interactions. Although field-level semantic knowledge may not be as rich as cumbersome instance-level semantic information, its conciseness and importance offer substantial potential for improving model performance.

Motivated by these analyses, we introduce  \underline{\textbf{LLaCTR}}, a novel \underline{\textbf{L}}ightweight \underline{\textbf{L}}LM-enh\underline{\textbf{a}}nced \underline{\textbf{CTR}} method through field-level enhancement, comprising two key components:

\begin{itemize}[topsep=0pt,leftmargin=10pt]
\setlength{\itemsep}{0pt}
\item \textbf{Self-supervised Field-feature Fine-tuning.} While LLMs are pre-trained on open-domain corpora, they may lack domain-specific knowledge relevant to recommender systems. This limitation can hinder the quality of the distilled field semantic knowledge. To address this, we design a self-supervised task where the LLM is prompted to predict the field to which a given feature belongs. This fine-tuning leverages rich domain knowledge encoded in feature-field relationships, enabling the LLM to better understand the field semantics. Notably, since the number of fields is small and fine-tuning only requires a limited sample of features, this approach is more efficient than directly extracting instance or user/item-level knowledge as in prior work. 
\item \textbf{Field Semantic-guided Enhancement.} 
We further utilize the field semantic knowledge (\ie field semantic embeddings) distilled from LLMs to enhance traditional CTR models in two key aspects: 1) field embeddings are utilized to guide the learning of feature embeddings via a specific alignment loss, injecting semantic knowledge into feature representations; 2) field embeddings are transformed into field interaction matrix through a dedicated network to supplement and enhance feature interaction modeling.
\end{itemize}

Notably, LLaCTR is both flexible and lightweight, allowing for seamless integration into existing CTR models as a plug-and-play enhancement. In our experiments, we incorporate LLaCTR into six conventional CTR methods and observe significant performance improvements across four real-world datasets (with an average increase of 2.24\%). Moreover, LLaCTR outperforms recent LLM-enhanced methods while incurring substantially lower computational overhead (by a factor of 10–100 times). 

Overall, this work makes the following contributions:

\begin{itemize}[leftmargin=*]
\item We identify the limitations of existing LLM-based CTR methods and advocate for leveraging field-level semantic knowledge to efficiently enhance traditional CTR models.
\item We propose a novel LLM-enhanced CTR method, LLaCTR, which employs self-supervised fine-tuning to distill high-quality field knowledge from LLMs and integrates this knowledge to improve both feature representation and feature interaction modeling.
\item We conduct extensive experiments to demonstrate that LLaCTR significantly improves the accuracy of traditional CTR models while being highly efficient and cost-effective.
\end{itemize}

\section{Preliminaries}
% In this section, we elaborate on the background of CTR prediction, and introduce the paradigm of current LLM-enhanced CTR models. 

% \subsection{Background of CTR Prediction}
\subsection{Task Formulation} 
This work focuses on click-through rate (CTR) prediction \cite{lin2023map, wang2022enhancing}, a core task in many personalized services (\eg recommender systems). Let $\mathcal{D}=\{(\mathbf{X}_i,y_i)\}_{i=1}^N$ denote the training dataset of user-item historical interaction records, where  $\mathbf{X}_i=[\mathbf{x}_{i1},\mathbf{x}_{i2},\ldots,\mathbf{x}_{iK}]$ represents the input features across  $K$-field for the $i$-th instance, and $\mathbf{x}_{ik}$ denotes the feature of the $k$-th field for the instance. In general, $\mathbf{x}_{ik}$ 
is a vector that can represent categorical, multi-valued, numerical features, \etc. The label  $y_i\in\{1,0\}$ indicates  whether the user has clicked on the item in this instance. The goal of CTR is to learn a model from $\mathcal{D}$ that can accurately predict the click probability of a new instance. 

\subsection{Traditional CTR Models} Traditional CTR models \cite{rendle2010factorization, guo2017deepfm, pan2018field, sun2021fm2, mao2023finalmlp, zhang2024wukong} primarily adopt deep learning paradigms and can generally be abstracted into three core components:

\textit{1) Feature Embedding Layer.} Given the high-dimensional and sparse nature of raw input features, feature embedding is commonly employed to map the raw features into dense representations:
\begin{equation}  
\mathbf{E}_i=\text{EmbeddingLayer}(\mathbf{X}_i),
\end{equation}
where $\mathbf{E}_i=[\mathbf{e}_{i1},\mathbf{e}_{i2},\ldots,\mathbf{e}_{iK}]\in\mathbb{R}^{K\times D}$ represents the learned feature embeddings of the instance $\mathbf{X}_i$, with $\mathbf{e}_{ik}$ denoting the D-dimensional embedding vector for $k$-th field feature. 

\textit{2) Feature Interaction Layer.} This crucial component is designed to effectively capture the complex feature interactions:
\begin{equation}  
\Phi(\mathbf{X}_i)=\text{FeatureInteraction}(\mathbf{X}_i, \mathbf{E}_i),
\end{equation}
where $\Phi(\cdot)$ is the hidden representation learned from feature interactions. Among various architectures, the most representative one is the factorization machine (FM) \cite{rendle2010factorization}, which explicitly model bi-level feature interactions: 
\begin{equation} 
\small
\Phi_{FM}(\mathbf{X}_i)={w_0} + \sum\limits_{k=1}^K {{\mathbf{w}_k}^T{\mathbf{x}_{ik}}}  + \sum\limits_{k=1}^K {\sum\limits_{l=k+1}^K \varsigma ({  { { \mathbf{x}_{ik} {\mathbf{x}_{il}}^T \text{<} \textbf{e}_{ik}, {\textbf{e}_{il}} \text{>}  } }}) },
\end{equation}
% \begin{equation}
% \begin{split}
% \Phi_{FM}(\mathbf{X}_i) &= w_0 + \sum\limits_{k=1}^K {\mathbf{w}_k}^T \mathbf{x}_{ik} \\
% &\quad + \sum\limits_{k=1}^K \sum\limits_{l=k+1}^K \varsigma \left( \mathbf{x}_{ik} \mathbf{x}_{il}^T \langle \mathbf{e}_{ik}, \mathbf{e}_{il} \rangle \right),
% \end{split}
% \end{equation}

where $w_0,\mathbf{w}_k$ denotes the learnable weights, and $<\cdot,\cdot>$ denotes the inner product of two vectors; $\varsigma(.)$ denotes the sum of all elements in the matrix. For convenience, 
here we adopt vector notation to represent the formula of Factorization Machines (FM), which is mathematically equivalent to the original formulation presented in \cite{rendle2010factorization}.
The features are explicitly bi-level interacted with $\mathbf{x}_{ik}{\mathbf{x}_{il}}^T$ and its contribution is controlled by their embeddings $\text{<}\mathbf{e}_{ik},\mathbf{e}_{il}\text{>}$.

Building on FM, recent works have explored various extensions \cite{xiao2017attentional, lu2021dual, tao2020hoafm}, such as introducing field-aware matrices \cite{juan2016field, pan2018field, sun2021fm2} or incorporating neural layers \cite{guo2017deepfm, zhang2019field}. Meanwhile, other studies have proposed novel architectures, such as convolution operator \cite{liu2019feature, li2019fi} and self-attention \cite{zhou2018deep, zhou2019deep}, to better capture complex feature interactions.

\textit{3) Prediction Layer.} Finally, the prediction layer transforms $\Phi(\mathbf{X}_i)$ into the model prediction $\hat{y_i}$:
\begin{equation}  
\hat{y_i}=\sigma[\text{PredictionLayer}(\Phi(\mathbf{X}_i))],
\end{equation}
where $\sigma(\cdot)$ is the sigmoid function. The model is typically optimized with the binary cross-entropy (BCE) loss \cite{mao2023cross} with:
\begin{equation}
\mathcal{L}_{BCE}(y,\hat{y})=-\frac{1}{N}\sum_{i=1}^N\left[y_i\log\left(\hat{y_i}\right)+(1-y_i)\log\left(1-\hat{y_i}\right)\right],
\end{equation}

While these traditional CTR models have made significant progress in the past few decades, they often suffer from the semantic information loss. Specifically, some methods rely solely on field IDs as categorical indicators \cite{juan2016field, pan2018field, sun2021fm2}, neglecting the rich semantic information in field descriptions. This limitation motivates our use of LLMs to enhance traditional CTR models with field-level semantic knowledge.

\begin{figure*}[t]
    \centering 
    \includegraphics[width=1\textwidth]{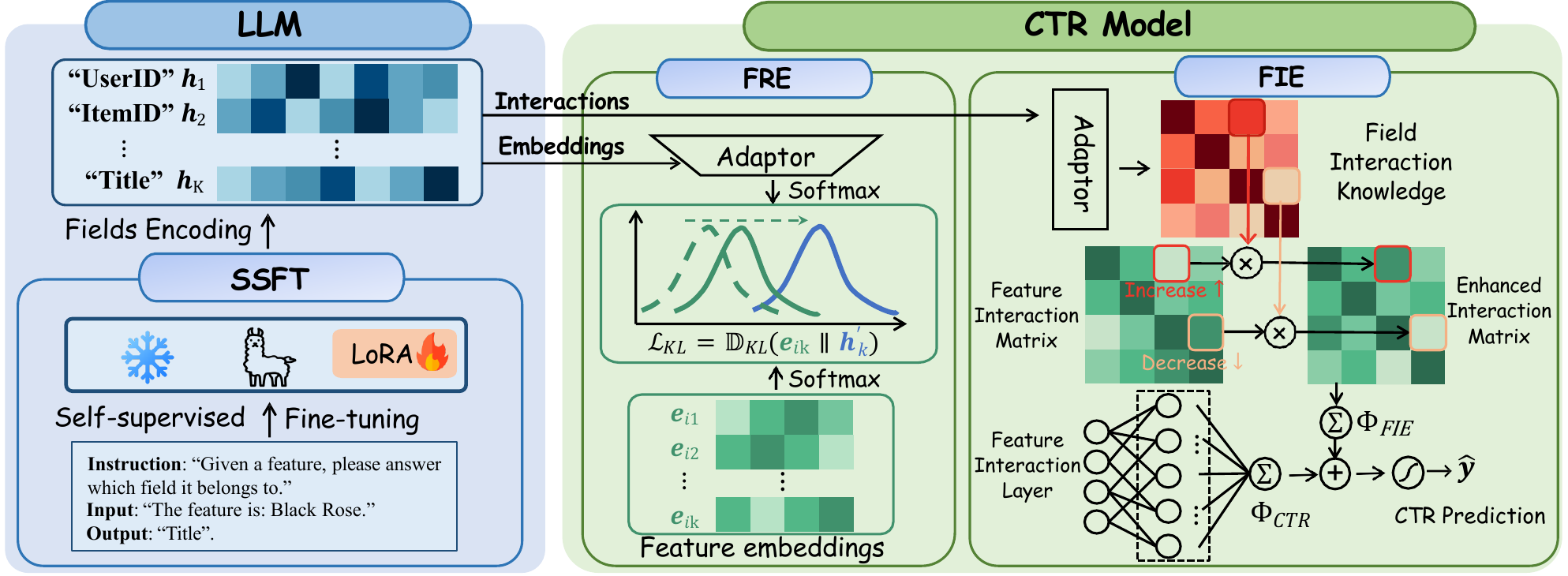}
    \vspace{-0.5cm}
    \caption{The overall framework of proposed LLaCTR.} 
    \label{fig:method} 
    \vspace{-0.3cm}
\end{figure*}

\subsection{LLM-enhanced CTR Models} To fully exploit semantic knowledge, LLMs have been extensively investigated for enhancing traditional CTR models. As illustrated in Figure~\ref{fig:field}, existing LLM-enhanced CTR methods primarily operate at the instance level or user/item level. Typically, these approaches first organize the features of  users, items, or instances into textual descriptions, which are then provided as prompts to the LLM for reasoning or summarization. The resulting semantic knowledge, often at the user/item or instance level, is further encoded into semantic embeddings to augment traditional CTR models. 

However, due to the large scale of users/items and instances in real-world applications, such strategies incur prohibitive computational costs during training, inference and encoding. For example, on representative datasets such as Amazon Video Games and MovieLens-1M, the total training time of existing LLM-enhanced CTR methods (e.g., KAR \cite{xi2024towards}, LLM-CF \cite{sun2024large}, CTRL \cite{li2023ctrl} and EASE \cite{qiu2024ease}) is over 290 times (on average) that of standard CTR models on average (as shown in Figure~\ref{fig:motivation_empirical}).
% (see Table~\ref{tab:Running-Time-Details} for details). 
Moreover, the complex and voluminous semantic knowledge generated at the user/item or instance level is often difficult for traditional CTR models to effectively assimilate and utilize. Given these limitations, it is imperative to explore new paradigms for integrating semantic knowledge into CTR models.

\section{Methodology}
\label{sec:Methodology}

In this section, we introduce LLaCTR, which enhances CTR prediction through field-level semantic knowledge from LLMs. As shown in Figure~\ref{fig:method}, LLaCTR first employs self-supervised field-feature fine-tuning to improve LLMs’ ability to capture field semantics (SSFT), and then leverages the field knowledge distilled from LLMs to enhance both feature representation (FRE) and feature interactions (FIE).

\subsection{Self-supervised Field-feature Fine-tuning (SSFT)} 
\label{sec:ssft}

LLMs are pre-trained on open-domain natural language corpora \cite{achiam2023gpt,dubey2024llama}, which often lack domain-specific knowledge relevant to CTR prediction. This limitation can hinder their ability to distill high-quality field semantic knowledge. Specifically, LLMs may not fully understand the meaning of field descriptions in CTR tasks or their relationships with features. To address this, we propose a \emph{self-supervised fine-tuning strategy} that prompts LLMs to predict the field to which a given feature belongs, thereby enhancing their understanding of field semantics.

\begin{figure}[t]
    \centering 
    \includegraphics[width=0.47\textwidth]{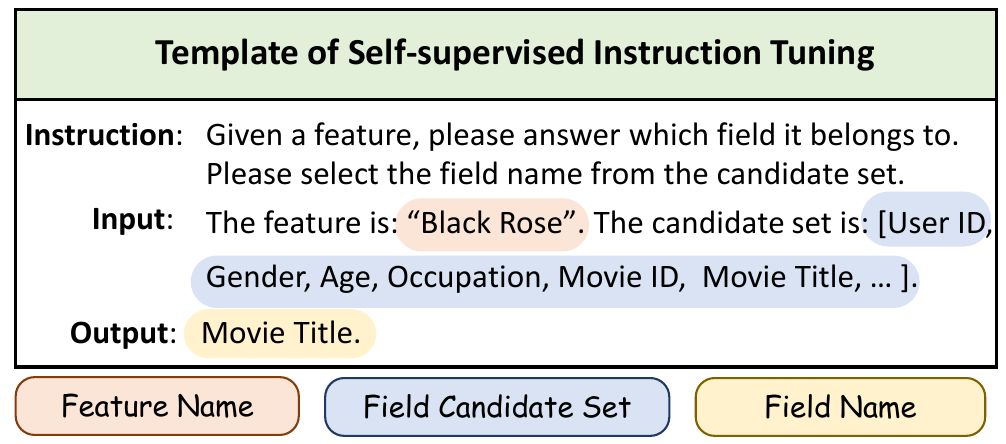}
    \vspace{-0.3cm}
    \caption{The template of self-supervised fine-tuning.} 
    \label{fig:InstructionPair} 
    \vspace{-0.5cm}
\end{figure}

Specifically, let $F_k$ represent the descriptions of the $k$-th field, and $f^k_j$ denote the description of the $j$-th feature within field $k$.  As shown in Figure~\ref{fig:InstructionPair}, we construct self-supervised training instances $(\mathcal P(f^k_j),F_k)$, where $\mathcal P(f^k_j)$ is a prompt containing the task description, feature descriptions, and candidate field descriptions, and $F_k$ denotes the $k$-th field description as the target response. These prompt-response pairs are used to fine-tune LLMs with language generative loss:
\begin{equation}
\mathcal{L}_{LG}(\mathcal P(f^k_j), F_k; \theta) = - \log P_{LLM} \left(F_k \mid \mathcal P(f^k_j)\right),
\end{equation}
where  $P_{LLM} \left(F_k \mid \mathcal P(f^k_j)\right)$ represents the LLM's probability of generating the correct field description $F_k$ given the prompt $\mathcal P(f^k_j)$. This loss improves the LLM's ability to capture field-feature correlations, injecting domain knowledge and enabling the generation of higher-quality semantic representations.

Besides the language generative loss, we find the contrastive loss \cite{he2020momentum} is also effective: 
\begin{equation}
\small
\mathcal{L}_{CL}(\mathcal P(f^k_j), F_k; \theta) = -\log\frac{\exp(cos(\mathcal{E}(\mathcal P(f_j^k)),\mathcal{E}(F_k))/\tau)}{\sum_{l}\exp(cos(\mathcal{E}(\mathcal P(f_j^k)),\mathcal{E}(F_l))/\tau)},
\end{equation}
where $cos(\cdot)$ denotes the cosine similarity function, $\tau$ is the temperature parameter and $\mathcal{E}(\cdot)$ are the LLM's encodings of the language descriptions, respectively. 

The contrastive loss is designed to align the semantic embeddings of prompts with their corresponding answers, thereby effectively injecting feature-field correlation knowledge in embeddings.  Since the contrastive loss operates directly on embeddings, which are subsequently used to enhance CTR models, this work prioritizes the use of contrastive loss and empirically find slightly better performance than generative loss (see Appendix~\ref{sec:appendix-Fine-tuning} for details).

After fine-tuning, the LLM's encodings of field descriptions are extracted:
\begin{equation}
\textbf{h}_k=\mathcal{E}(F_k).
\end{equation}
For convenience, we collect the field embeddings as a matrix $\textbf{H}=\{\textbf{h}_k\}_{k=1}^K$. These field embeddings will further utilized to enhance traditional CTR models.

% \begin{wrapfigure}{r}{0.5\textwidth} % 图片靠右，宽度为页面宽度的 50%
%     \vspace{-10pt} % 调整图片与上方文字的间距
%     \centering
%     \includegraphics[width=0.5\textwidth]{fig/template25.pdf}
%     \vspace{-0.4cm} % 调整图片与下方文字的间距
%     \caption{The template of self-supervised fine-tuning.}
%     \label{fig:InstructionPair}
%     \vspace{-0.3cm} % 调整图片与下方文字的间距
% \end{wrapfigure}

Notably, such field-level operations are highly efficient, as the LLMs require to process only a small number of fields (typically less than 100) rather than millions of instances or user/item pairs in the experimental datasets. 
% This difference in scale is even more pronounced in real-world industrial recommender systems, where there may be thousands of fields, but hundreds of millions of users/items and billions of instances. 
Some readers may be concerned about the potential inefficiency of self-supervised fine-tuning. However, this process is also lightweight, as it only requires sampling a small number of features per field for training. For example, sampling just 500 features per field is sufficient to achieve significantly better performance compared to recent LLM-enhanced methods, while incurring substantially lower training time. Empirical comparisons of efficiency can be found in Figure~\ref{fig:expxx}. The detailed time efficiency analyses of LLaCTR and other LLM-enhanced baselines are provided in Section~\ref{sec:efficiency-study}.

\subsection{Feature Representation Enhancement (FRE)} 
\label{sec:fre}

The quality of feature representations is critical for the success of CTR models. However, existing methods often overlook the semantic knowledge contained in field descriptions \cite{lin2024clickprompt, xi2024towards}. To address this, we transfer the field semantic knowledge distilled from LLMs to enhance feature embeddings.

Given the embedding space gap between LLMs and CTR models, we first map original field embeddings $h_k$ into the embedding space of CTR models using a learnable adaptor:
\begin{equation}
\textbf{h}^{'}_k= \text{Adaptor}(\textbf{h}_k),
\end{equation}
where $\text{Adaptor}(\cdot)$ can be implemented as a simple linear transformation. We then align feature embeddings with their corresponding field embeddings through normalized KL-divergence \cite{tian2019contrastive}:
\begin{equation}
\begin{aligned}
\mathbf{h}^{'}_k \leftarrow \text{Softmax}(\mathbf{h}^{'}_k), \
\mathbf{e}_{ik} \leftarrow \text{Softmax}(\mathbf{e}_{ik}),
\end{aligned}
\end{equation}
\begin{equation}
\label{equ:cl}
\small
\mathcal{L}_{KL}=\sum_{k=1}^K\sum_{i=1}^N\mathbb{D}_{\text{KL}}(\mathbf{e}_{ik},\mathbf{h}^{'}_k)
= \sum_{k=1}^K\sum_{i=1}^N \mathbf{e}_{ik}\cdot \ln({\mathbf{e}_{ik}}/{\mathbf{h}_{k}^{'}}).
\end{equation}
Specifically, here we employ normalized KL-divergence, as it eliminates the influence of embedding magnitude and focuses on aligning the underlying distributions. We also experimented with other commonly used alignment losses, such as contrastive loss, but found that normalized KL-divergence consistently yields superior performance (see Appendix~\ref{sec:appendix-Alignment} for details). By minimizing $\mathcal{L}_{KL}$, feature embeddings are able to absorb semantic knowledge from the field embeddings, resulting in higher-quality representations. Intuitively, the field semantic embeddings serve as prototypes, encouraging the feature embeddings close to these semantic centers during training. 

The overall training objective combines semantic alignment with the original CTR objective:
\begin{equation}
\mathcal{L}= \mathcal{L}_{BCE} + \lambda_{kl}\mathcal{L}_{KL},
\end{equation}
where hyperparameter $\lambda_{kl}$ balances their contributions.

\subsection{Feature Interaction Enhancement (FIE)} 
\label{sec:fie}

Field information has been shown beneficial to improve feature interaction modeling in CTR tasks \cite{juan2016field, rendle2010factorization, sun2021fm2}. However, these traditional method only utilize field ID information by employing implicit embedding layers \cite{juan2016field} or learnable parameters matrices \cite{rendle2010factorization, sun2021fm2}, without taking the textual information of fields into account. Field semantic information provides new insights into feature interactions. For example, semantically, interactions between the features belonging to "user income" and "item price" are likely highly useful for click prediction.

To leverage this insight, we first compute the importance of feature interactions of two fields based on their semantic embeddings:
\begin{equation}
m_{ij}^F=\text{FieldInteraction}(\mathbf{h}_i, \mathbf{h}_j),
\end{equation}
where $\text{FieldInteraction}(\cdot)$ denotes the field interaction layer, which can be implemented by various architecture. In practice, we find that cosine similarity is effective --- intuitively, the fields with similar semantic may provide more insights on prediction. For convenience, we collect all pairwise field interaction scores into a matrix $\mathbf{M}^F$. Subsequently, we employ a learnable adaptor to re-scale these importance scores, \ie  $\mathbf{M}^{'} = \text{Adaptor}(\mathbf{M}^F)$, where the adaptor can be simply implemented via a linear layer.

The field-aware importance scores derived from semantic embeddings are then utilized to guide the learning of feature interactions. Specifically, we explicitly incorporate these scores into a bi-level feature interaction modeling, thereby enhancing the feature interaction layer of CTR methods:
\begin{equation} 
\small
\Phi^{new}(\mathbf{X}_i) = \Phi(\mathbf{X}_i) + \lambda_{fm}\cdot \sum\limits_{k=1}^K {\sum\limits_{l=k+1}^K \varsigma ({  { { \mathbf{x}_{ik} {\mathbf{x}_{il}}^T \text{<} \textbf{e}_{ik}, {\textbf{e}_{il}}^T \text{>}  m_{kl}^{'}} }}) }  ,
\end{equation} 

where the field-aware importance score  $m_{kl}^{'}$ modulates the strength of feature interactions. Here, we augment the original feature interaction layer by integrating an additional field-aware feature interaction component, with $\lambda_{fm}$ controlling the relative contribution of the two components.

Notably, our proposed feature interaction enhancement strategy can be seamlessly integrated into a wide range of existing CTR models. This generality motivates our approach of introducing a supplementary field-aware feature interaction component, rather than modifying the inherent feature interaction architecture of each model. In our experiments, we incorporate our LLaCTR module into six representative CTR methods, and observe substantial performance improvements across almost all cases. Detailed results are presented in Section~\ref{sec:experiments}, and the implementation details are provided in the Appendix~\ref{sec:appendix-experiments}.

\begin{table}[t]
\centering
\caption{Statistics of the datasets.}
\vspace{-0.3cm}
\label{tab:datasets}
\scalebox{0.85}{
\begin{tabular}{c|cccc}
\Xhline{1.2pt}
\hline
Dataset & Gift Cards & Video Games & Digital Music & MovieLens-1M \bigstrut \\ \hline
\#Field & 13 & 13 & 13 & 8 \bigstrut[t]\\
\#Feature & 1,981,330 & 60,119,995 & 1,695,642 & 9,001,881 \\
\#User & 132,732 & 2,766,656 & 100,952 & 6,040 \\
\#Item & 1,137 & 137,249 & 70,511 & 3,706 \\
\#Interaction & 152,410 & 4,624,615 & 130,434 & 1,000,209 \\
Density & $1.01\times10^{-3}$ & $1.22\times10^{-5}$ & $1.83\times10^{-5}$ & $4.47\times10^{-2}$ \bigstrut[b]\\ \hline
\Xhline{1.2pt}
\end{tabular}
}
\vspace{-0.3cm}
\end{table}

\begin{table*}[t]
\centering
\caption{Performance comparisons of LLaCTR with existing PLM/LLM-enhanced methods. The results of PLM-enhanced methods are reported on their best backbones. The best result is bolded and the blue-colored zone indicates that LLaCTR is better than the basic CTR backbone.  The mark "*" indicates the improvement is statistically significant ($p$-value $< 0.05$).}
\vspace{-0.3cm}
\label{tab:performance}
\scalebox{0.85}{
\begin{tabular}{cccccccccccccc}
\Xhline{1.2pt}
\hline
\multicolumn{2}{c|}{} & \multicolumn{3}{c|}{Gift Cards} & \multicolumn{3}{c|}{Video Games} & \multicolumn{3}{c|}{Digital Music} & \multicolumn{3}{c}{MovieLens-1M} \bigstrut \\ \cline{3-14} 
\multicolumn{2}{c|}{\multirow{-2}{*}{Method}} & Logloss ↓ & AUC ↑ & \multicolumn{1}{c|}{RelaImpr} & Logloss ↓ & AUC ↑ & \multicolumn{1}{c|}{RelaImpr} & Logloss ↓ & AUC ↑ & \multicolumn{1}{c|}{RelaImpr} & Logloss ↓ & AUC ↑ & RelaImpr \bigstrut \\ \hline
% \multicolumn{14}{c}{PLM-enhanced CTR Methods} \bigstrut \\ \hline
\multicolumn{2}{c|}{CELA} & 0.4658 & 0.6746 & \multicolumn{1}{c|}{1.02\%} & 0.5798 & 0.6775 & \multicolumn{1}{c|}{0.44\%} & 0.3364 & 0.7638 & \multicolumn{1}{c|}{0.37\%} & 0.4031 & 0.8373 & 0.24\% \bigstrut[t] \\
\multicolumn{2}{c|}{ClickPrompt} & 0.4656 & 0.6754 & \multicolumn{1}{c|}{1.48\%} & 0.5747 & 0.6799 & \multicolumn{1}{c|}{1.79\%} & 0.3356 & 0.7652 & \multicolumn{1}{c|}{0.90\%} & 0.4025 & 0.8379 & 0.42\% \bigstrut[b] \\ \hline
% \multicolumn{14}{c}{LLM-enhanced CTR Methods} \bigstrut \\ \hline
\multicolumn{1}{c|}{} & \multicolumn{1}{c|}{Base} & 0.4606 & 0.6621 & \multicolumn{1}{c|}{} & 0.5805 & 0.6673 & \multicolumn{1}{c|}{} & 0.3359 & 0.7612 & \multicolumn{1}{c|}{} & 0.4081 & 0.8354 &  \bigstrut[t] \\
\multicolumn{1}{c|}{} & \multicolumn{1}{c|}{KAR} & 0.4551 & 0.6643 & \multicolumn{1}{c|}{1.34\%} & 0.5803 & 0.6705 & \multicolumn{1}{c|}{1.93\%} & 0.3410 & 0.7607 & \multicolumn{1}{c|}{-0.21\%} & 0.4073 & 0.8358 & 0.13\% \\
\multicolumn{1}{c|}{} & \multicolumn{1}{c|}{LLM-CF} & 0.4580 & 0.6666 & \multicolumn{1}{c|}{2.77\%} & 0.5805 & 0.6682 & \multicolumn{1}{c|}{0.58\%} & \textbf{0.3351} & \textbf{0.7633} & \multicolumn{1}{c|}{\textbf{0.78\%}} & 0.4128 & 0.8337 & -0.50\% \\
\multicolumn{1}{c|}{} & \multicolumn{1}{c|}{CTRL} & 0.4595 & 0.6636 & \multicolumn{1}{c|}{0.93\%} & 0.5802 & 0.6708 & \multicolumn{1}{c|}{2.13\%} & 0.3360 & 0.7611 & \multicolumn{1}{c|}{-0.04\%} & 0.4128 & 0.8336 & -0.53\% \\
\multicolumn{1}{c|}{} & \multicolumn{1}{c|}{EASE} & 0.4534 & 0.6667 & \multicolumn{1}{c|}{2.84\%} & 0.5804 & 0.6701 & \multicolumn{1}{c|}{1.69\%} & 0.3359 & 0.7625 & \multicolumn{1}{c|}{0.48\%} & 0.4101 & 0.8357 & 0.10\% \\
\multicolumn{1}{c|}{\multirow{-6}{*}{FM}} & \multicolumn{1}{c|}{LLaCTR} & \cellcolor[HTML]{DDEBF7}\textbf{0.4500} & \cellcolor[HTML]{DDEBF7}\textbf{0.6673} & \multicolumn{1}{c|}{\cellcolor[HTML]{DDEBF7}\textbf{3.21\%*}} & \cellcolor[HTML]{DDEBF7}\textbf{0.5801} & \cellcolor[HTML]{DDEBF7}\textbf{0.6718} & \multicolumn{1}{c|}{\cellcolor[HTML]{DDEBF7}\textbf{2.70\%*}} & 0.3362 & \cellcolor[HTML]{DDEBF7}0.7621 & \multicolumn{1}{c|}{\cellcolor[HTML]{DDEBF7}0.32\%} & \cellcolor[HTML]{DDEBF7}\textbf{0.4050} & \cellcolor[HTML]{DDEBF7}\textbf{0.8364} & \cellcolor[HTML]{DDEBF7}\textbf{0.30\%*} \bigstrut[b] \\ \hline
\multicolumn{1}{c|}{} & \multicolumn{1}{c|}{Base} & 0.4751 & 0.6713 & \multicolumn{1}{c|}{} & 0.5762 & 0.6692 & \multicolumn{1}{c|}{} & 0.3478 & 0.7582 & \multicolumn{1}{c|}{} & 0.4190 & 0.8342 &  \bigstrut[t] \\
\multicolumn{1}{c|}{} & \multicolumn{1}{c|}{KAR} & 0.4753 & 0.6778 & \multicolumn{1}{c|}{3.80\%} & 0.5871 & 0.6720 & \multicolumn{1}{c|}{1.65\%} & 0.3439 & 0.7587 & \multicolumn{1}{c|}{0.18\%} & 0.4207 & 0.8335 & -0.20\% \\
\multicolumn{1}{c|}{} & \multicolumn{1}{c|}{LLM-CF} & 0.4723 & 0.6736 & \multicolumn{1}{c|}{1.36\%} & 0.5769 & 0.6730 & \multicolumn{1}{c|}{2.28\%} & 0.3441 & 0.7588 & \multicolumn{1}{c|}{0.21\%} & 0.4140 & 0.8353 & 0.34\% \\
\multicolumn{1}{c|}{} & \multicolumn{1}{c|}{CTRL} & 0.4746 & 0.6744 & \multicolumn{1}{c|}{1.83\%} & \textbf{0.5757} & 0.6758 & \multicolumn{1}{c|}{3.92\%} & 0.3432 & 0.7576 & \multicolumn{1}{c|}{-0.26\%} & 0.4142 & 0.8352 & 0.31\% \\
\multicolumn{1}{c|}{} & \multicolumn{1}{c|}{EASE} & 0.4748 & 0.6739 & \multicolumn{1}{c|}{1.52\%} & 0.5807 & 0.6738 & \multicolumn{1}{c|}{2.74\%} & 0.3360 & 0.7604 & \multicolumn{1}{c|}{0.83\%} & 0.4180 & 0.8349 & 0.22\% \\
\multicolumn{1}{c|}{\multirow{-6}{*}{DeepFM}} & \multicolumn{1}{c|}{LLaCTR} & \cellcolor[HTML]{DDEBF7}\textbf{0.4657} & \cellcolor[HTML]{DDEBF7}\textbf{0.6797} & \multicolumn{1}{c|}{\cellcolor[HTML]{DDEBF7}\textbf{4.90\%*}} & 0.5834 & \cellcolor[HTML]{DDEBF7}\textbf{0.6744} & \multicolumn{1}{c|}{\cellcolor[HTML]{DDEBF7}\textbf{3.10\%*}} & \cellcolor[HTML]{DDEBF7}\textbf{0.3350} & \cellcolor[HTML]{DDEBF7}\textbf{0.7687} & \multicolumn{1}{c|}{\cellcolor[HTML]{DDEBF7}\textbf{4.06\%*}} & \cellcolor[HTML]{DDEBF7}\textbf{0.4049} & \cellcolor[HTML]{DDEBF7}\textbf{0.8354} & \cellcolor[HTML]{DDEBF7}\textbf{0.38\%*} \bigstrut[b] \\ \hline
\multicolumn{1}{c|}{} & \multicolumn{1}{c|}{Base} & 0.4664 & 0.6691 & \multicolumn{1}{c|}{} & 0.5762 & 0.6683 & \multicolumn{1}{c|}{} & 0.3420 & 0.7553 & \multicolumn{1}{c|}{} & 0.4017 & 0.8365 &  \bigstrut[t] \\
\multicolumn{1}{c|}{} & \multicolumn{1}{c|}{KAR} & 0.4661 & 0.6694 & \multicolumn{1}{c|}{0.19\%} & 0.5791 & 0.6741 & \multicolumn{1}{c|}{3.42\%} & 0.3432 & 0.7575 & \multicolumn{1}{c|}{0.88\%} & 0.4020 & 0.8360 & -0.15\% \\
\multicolumn{1}{c|}{} & \multicolumn{1}{c|}{LLM-CF} & 0.4672 & 0.6698 & \multicolumn{1}{c|}{0.47\%} & 0.5806 & 0.6702 & \multicolumn{1}{c|}{1.13\%} & 0.3397 & 0.7567 & \multicolumn{1}{c|}{0.57\%} & 0.4044 & 0.8371 & 0.18\% \\
\multicolumn{1}{c|}{} & \multicolumn{1}{c|}{CTRL} & 0.4679 & 0.6699 & \multicolumn{1}{c|}{0.50\%} & \textbf{0.5776} & 0.6755 & \multicolumn{1}{c|}{4.28\%} & 0.3414 & 0.7561 & \multicolumn{1}{c|}{0.33\%} & \textbf{0.4004} & \textbf{0.8372} & \textbf{0.21\%} \\
\multicolumn{1}{c|}{} & \multicolumn{1}{c|}{EASE} & 0.4661 & 0.6703 & \multicolumn{1}{c|}{0.74\%} & 0.5795 & 0.6765 & \multicolumn{1}{c|}{4.84\%} & 0.3415 & 0.7571 & \multicolumn{1}{c|}{0.71\%} & 0.4030 & 0.8369 & 0.12\% \\
\multicolumn{1}{c|}{\multirow{-6}{*}{FwFM}} & \multicolumn{1}{c|}{LLaCTR} & \cellcolor[HTML]{DDEBF7}\textbf{0.4659} & \cellcolor[HTML]{DDEBF7}\textbf{0.6718} & \multicolumn{1}{c|}{\cellcolor[HTML]{DDEBF7}\textbf{1.61\%*}} & 0.5827 & \cellcolor[HTML]{DDEBF7}\textbf{0.6780} & \multicolumn{1}{c|}{\cellcolor[HTML]{DDEBF7}\textbf{5.73\%*}} & \cellcolor[HTML]{DDEBF7}\textbf{0.3362} & \cellcolor[HTML]{DDEBF7}\textbf{0.7585} & \multicolumn{1}{c|}{\cellcolor[HTML]{DDEBF7}\textbf{1.27\%*}} & \cellcolor[HTML]{DDEBF7}0.4014 & \cellcolor[HTML]{DDEBF7}0.8367 & \cellcolor[HTML]{DDEBF7}0.05\% \bigstrut[b] \\ \hline
\multicolumn{1}{c|}{} & \multicolumn{1}{c|}{Base} & 0.4579 & 0.6747 & \multicolumn{1}{c|}{} & 0.5824 & 0.6706 & \multicolumn{1}{c|}{} & 0.3315 & 0.7628 & \multicolumn{1}{c|}{} & 0.4077 & 0.8325 &  \bigstrut[t] \\
\multicolumn{1}{c|}{} & \multicolumn{1}{c|}{KAR} & 0.4595 & 0.6754 & \multicolumn{1}{c|}{0.42\%} & 0.5767 & 0.6711 & \multicolumn{1}{c|}{0.29\%} & 0.3347 & 0.7640 & \multicolumn{1}{c|}{0.43\%} & 0.4062 & 0.8332 & 0.21\% \\
\multicolumn{1}{c|}{} & \multicolumn{1}{c|}{LLM-CF} & 0.4652 & 0.6740 & \multicolumn{1}{c|}{-0.43\%} & 0.5877 & 0.6709 & \multicolumn{1}{c|}{0.20\%} & 0.3317 & 0.7627 & \multicolumn{1}{c|}{-0.04\%} & 0.4066 & 0.8329 & 0.11\% \\
\multicolumn{1}{c|}{} & \multicolumn{1}{c|}{CTRL} & 0.4593 & 0.6751 & \multicolumn{1}{c|}{0.21\%} & 0.5826 & 0.6714 & \multicolumn{1}{c|}{0.47\%} & \textbf{0.3310} & \textbf{0.7650} & \multicolumn{1}{c|}{\textbf{0.82\%}} & 0.4059 & 0.8332 & 0.21\% \\
\multicolumn{1}{c|}{} & \multicolumn{1}{c|}{EASE} & 0.4590 & 0.6753 & \multicolumn{1}{c|}{0.33\%} & 0.5750 & 0.6729 & \multicolumn{1}{c|}{1.35\%} & 0.3311 & 0.7637 & \multicolumn{1}{c|}{0.33\%} & 0.4056 & 0.8334 & 0.27\% \\
\multicolumn{1}{c|}{\multirow{-6}{*}{FmFM}} & \multicolumn{1}{c|}{LLaCTR} & \cellcolor[HTML]{DDEBF7}\textbf{0.4560} & \cellcolor[HTML]{DDEBF7}\textbf{0.6761} & \multicolumn{1}{c|}{\cellcolor[HTML]{DDEBF7}\textbf{0.78\%*}} & \cellcolor[HTML]{DDEBF7}\textbf{0.5725} & \cellcolor[HTML]{DDEBF7}\textbf{0.6765} & \multicolumn{1}{c|}{\cellcolor[HTML]{DDEBF7}\textbf{3.45\%*}} & 0.3323 & 0.7615 & \multicolumn{1}{c|}{-0.50\%} & \cellcolor[HTML]{DDEBF7}\textbf{0.4055} & \cellcolor[HTML]{DDEBF7}\textbf{0.8335} & \cellcolor[HTML]{DDEBF7}\textbf{0.30\%*} \bigstrut[b] \\ \hline
\multicolumn{1}{c|}{} & \multicolumn{1}{c|}{Base} & 0.4774 & 0.6720 & \multicolumn{1}{c|}{} & 0.5830 & 0.6761 & \multicolumn{1}{c|}{} & 0.3423 & 0.7557 & \multicolumn{1}{c|}{} & 0.4086 & 0.8301 &  \bigstrut[t] \\
\multicolumn{1}{c|}{} & \multicolumn{1}{c|}{KAR} & 0.4881 & 0.6726 & \multicolumn{1}{c|}{0.35\%} & 0.5820 & 0.6817 & \multicolumn{1}{c|}{3.19\%} & 0.3429 & 0.7546 & \multicolumn{1}{c|}{-0.41\%} & 0.4094 & 0.8290 & -0.33\% \\
\multicolumn{1}{c|}{} & \multicolumn{1}{c|}{LLM-CF} & 0.4781 & 0.6731 & \multicolumn{1}{c|}{0.62\%} & 0.5845 & 0.6785 & \multicolumn{1}{c|}{1.36\%} & 0.3369 & 0.7560 & \multicolumn{1}{c|}{0.13\%} & 0.4041 & 0.8299 & -0.06\% \\
\multicolumn{1}{c|}{} & \multicolumn{1}{c|}{CTRL} & 0.4719 & 0.6781 & \multicolumn{1}{c|}{3.55\%} & 0.5835 & 0.6790 & \multicolumn{1}{c|}{1.65\%} & \textbf{0.3399} & \textbf{0.7561} & \multicolumn{1}{c|}{\textbf{0.15\%}} & 0.4033 & 0.8328 & 0.82\% \\
\multicolumn{1}{c|}{} & \multicolumn{1}{c|}{EASE} & 0.4773 & 0.6736 & \multicolumn{1}{c|}{0.93\%} & 0.5776 & 0.6796 & \multicolumn{1}{c|}{1.97\%} & 0.3414 & 0.7560 & \multicolumn{1}{c|}{0.12\%} & 0.4049 & 0.8309 & 0.24\% \\
\multicolumn{1}{c|}{\multirow{-6}{*}{FinalMLP}} & \multicolumn{1}{c|}{LLaCTR} & \cellcolor[HTML]{DDEBF7}\textbf{0.4717} & \cellcolor[HTML]{DDEBF7}\textbf{0.6818} & \multicolumn{1}{c|}{\cellcolor[HTML]{DDEBF7}\textbf{5.71\%*}} & \cellcolor[HTML]{DDEBF7}\textbf{0.5738} & \cellcolor[HTML]{DDEBF7}\textbf{0.6832} & \multicolumn{1}{c|}{\cellcolor[HTML]{DDEBF7}\textbf{4.02\%*}} & \cellcolor[HTML]{DDEBF7}0.3411 & 0.7552 & \multicolumn{1}{c|}{-0.19\%} & \cellcolor[HTML]{DDEBF7}\textbf{0.4009} & \cellcolor[HTML]{DDEBF7}\textbf{0.8340} & \cellcolor[HTML]{DDEBF7}\textbf{1.18\%*} \bigstrut[b] \\ \hline
\multicolumn{1}{c|}{} & \multicolumn{1}{c|}{Base} & 0.4755 & 0.6728 & \multicolumn{1}{c|}{} & 0.5819 & 0.6767 & \multicolumn{1}{c|}{} & 0.3402 & 0.7514 & \multicolumn{1}{c|}{} & 0.4088 & 0.8323 &  \bigstrut[t] \\
\multicolumn{1}{c|}{} & \multicolumn{1}{c|}{KAR} & 0.4794 & 0.6721 & \multicolumn{1}{c|}{-0.43\%} & 0.5927 & 0.6758 & \multicolumn{1}{c|}{-0.53\%} & 0.3429 & 0.7506 & \multicolumn{1}{c|}{-0.31\%} & 0.4099 & 0.8316 & -0.21\% \\
\multicolumn{1}{c|}{} & \multicolumn{1}{c|}{LLM-CF} & 0.4756 & 0.6752 & \multicolumn{1}{c|}{1.36\%} & 0.5805 & 0.6788 & \multicolumn{1}{c|}{1.15\%} & 0.3379 & 0.7560 & \multicolumn{1}{c|}{1.83\%} & 0.4084 & 0.8293 & -0.90\% \\
\multicolumn{1}{c|}{} & \multicolumn{1}{c|}{CTRL} & 0.4753 & 0.6769 & \multicolumn{1}{c|}{2.37\%} & 0.5792 & 0.6800 & \multicolumn{1}{c|}{1.84\%} & 0.3399 & 0.7561 & \multicolumn{1}{c|}{1.85\%} & 0.4029 & 0.8367 & 1.31\% \\
\multicolumn{1}{c|}{} & \multicolumn{1}{c|}{EASE} & 0.4748 & 0.6758 & \multicolumn{1}{c|}{1.71\%} & 0.5783 & 0.6801 & \multicolumn{1}{c|}{1.91\%} & 0.3381 & 0.7556 & \multicolumn{1}{c|}{1.67\%} & 0.4031 & 0.8334 & 0.33\% \\
\multicolumn{1}{c|}{\multirow{-6}{*}{WuKong}} & \multicolumn{1}{c|}{LLaCTR} & \cellcolor[HTML]{DDEBF7}\textbf{0.4745} & \cellcolor[HTML]{DDEBF7}\textbf{0.6802} & \multicolumn{1}{c|}{\cellcolor[HTML]{DDEBF7}\textbf{4.24\%*}} & \cellcolor[HTML]{DDEBF7}\textbf{0.5763} & \cellcolor[HTML]{DDEBF7}\textbf{0.6819} & \multicolumn{1}{c|}{\cellcolor[HTML]{DDEBF7}\textbf{2.91\%*}} & \cellcolor[HTML]{DDEBF7}\textbf{0.3377} & \cellcolor[HTML]{DDEBF7}\textbf{0.7564} & \multicolumn{1}{c|}{\cellcolor[HTML]{DDEBF7}\textbf{1.99\%*}} & \cellcolor[HTML]{DDEBF7}\textbf{0.4027} & \cellcolor[HTML]{DDEBF7}\textbf{0.8397} & \cellcolor[HTML]{DDEBF7}\textbf{2.21\%*} \bigstrut[b] \\ \hline
\multicolumn{2}{c|}{\textbf{Average RelaImpr}} & \multicolumn{3}{c|}{{\color[HTML]{FF0000} \textbf{3.41\%}}} & \multicolumn{3}{c|}{{\color[HTML]{FF0000} \textbf{3.65\%}}} & \multicolumn{3}{c|}{{\color[HTML]{FF0000} \textbf{1.16\%}}} & \multicolumn{3}{c}{{\color[HTML]{FF0000} \textbf{0.74\%}}} \bigstrut \\ \hline
\Xhline{1.2pt}
\end{tabular}
}
\vspace{-0.1cm}
\end{table*}

\section{Experiments}
\label{sec:experiments}
We aim to answer the following research questions:
\begin{itemize}[leftmargin=*]
  \item $\mathbf{RQ1:}$ How does LLaCTR perform compared with the state-of-the-art CTR methods?
  \item $\mathbf{RQ2:}$ How is the efficiency of LLaCTR compared with existing LLM-enhanced CTR methods?
  \item $\mathbf{RQ3:}$ What are the impacts of different components of LLaCTR?
\end{itemize}

\subsection{Experimental Setup}

\begin{figure*}[t]
    \centering
    \vspace{-0.1cm}
    \begin{subfigure}[b]{0.495\textwidth}
        \includegraphics[width=\textwidth]{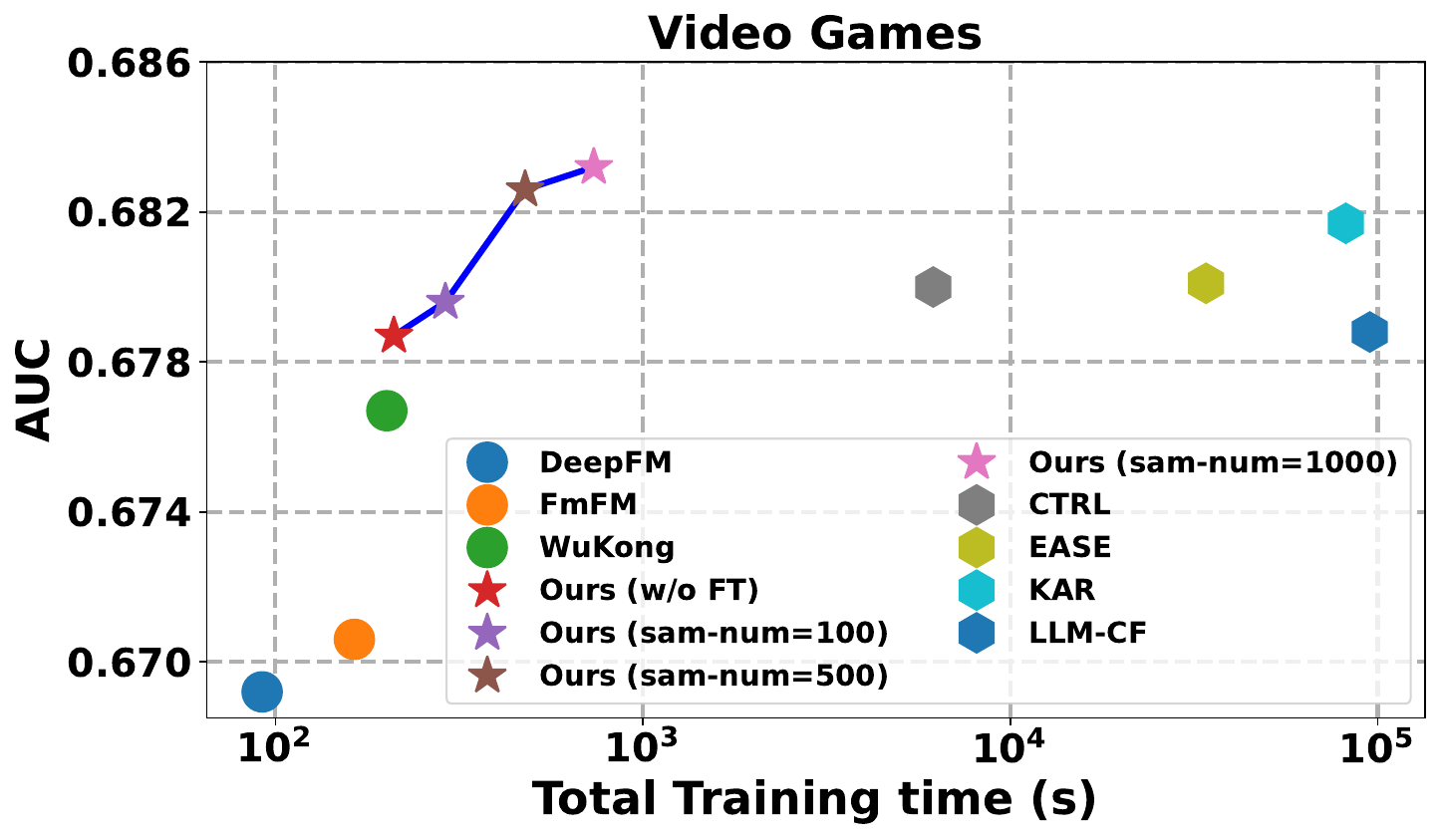}
        % \caption{Video Games}
    \end{subfigure}
    \begin{subfigure}[b]{0.495\textwidth}
        \includegraphics[width=\textwidth]{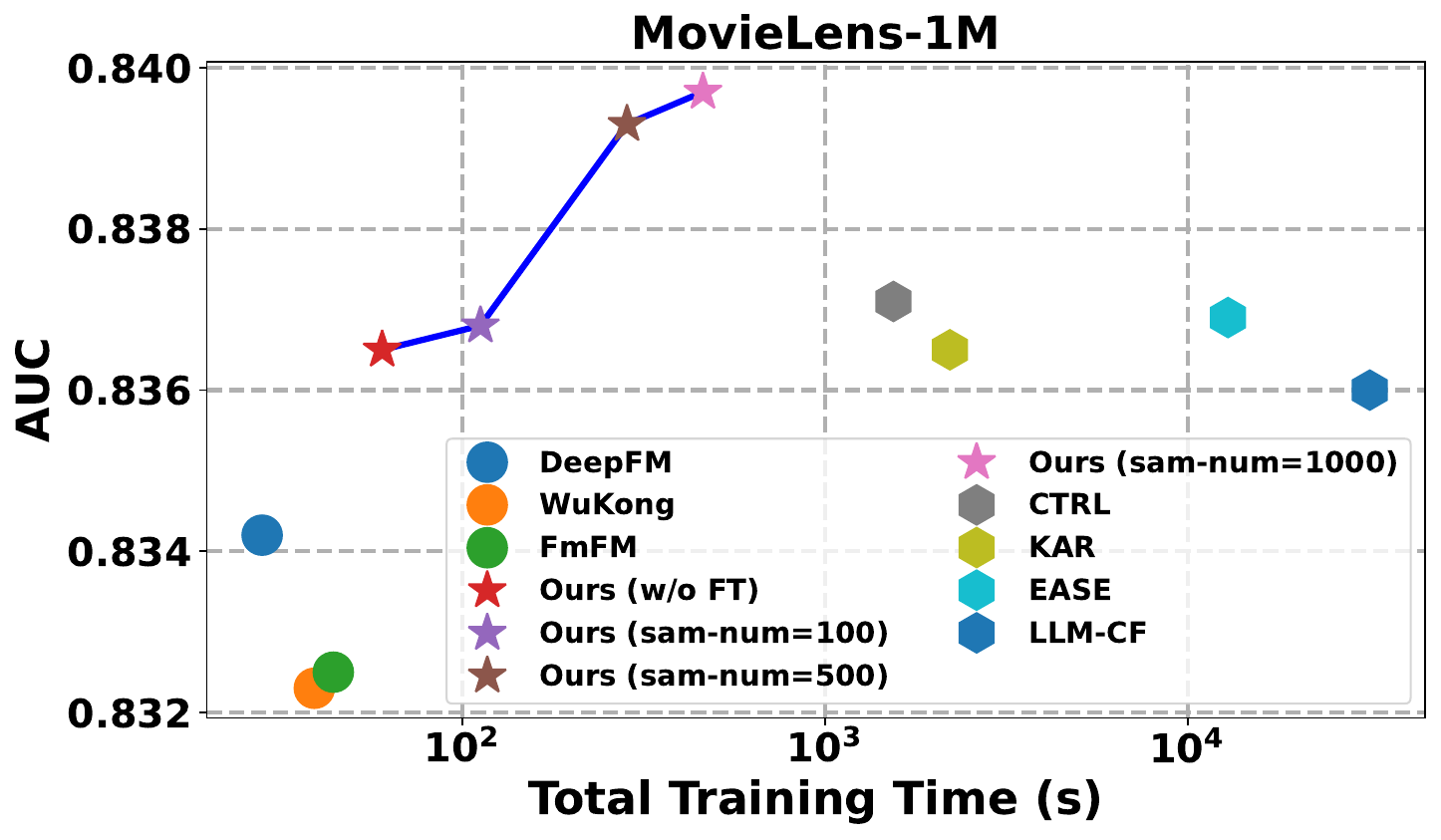}
        % \caption{MovieLens-1M}
    \end{subfigure}
    \vspace{-0.3cm}
    \caption{AUC and training time of compared methods. The “sam-num” is the sampling feature number of each field, and "w/o FT" represents fine-tuning has been removed.} 
    \label{fig:expxx}
    \vspace{-0.1cm}
\end{figure*}

\subsubsection{Datasets}
Following previous work \cite{lin2024clickprompt, xi2024towards}, we conduct experiments on four widely used public datasets in LLM-enhanced CTR task: \textit{Amazon Gift Cards}, \textit{Amazon Video Games}, \textit{Amazon Digital Music} and \textit{MovieLens-1M}. 
% Amazon Review Datasets\footnote{\url{https://amazon-reviews-2023.github.io/}} are well-known e-commercial datasets from the Amazon website with ratings ranging from 1 to 5. We take the items with a rating of greater than 3 as positive and the rest as negative \cite{li2023ctrl}.
% MovieLens-1M Dataset\footnote{\url{https://grouplens.org/datasets/MovieLens/1m/}} is a benchmark recommendation dataset from the Movielens website with ratings ranging from 1 to 5. We binarize the ratings with a threshold of 4, while removing neutral samples with ratings equal to 3 \cite{lin2024clickprompt}.
\begin{itemize}[topsep=0pt,leftmargin=10pt]
    \setlength{\itemsep}{0pt}
    \item \textbf{Amazon Review Datasets}\footnote{\url{https://amazon-reviews-2023.github.io/}} are well-known e-commercial datasets \cite{sun2024large, li2023ctrl, lin2024clickprompt, bao2023bi} with ratings ranging from 1 to 5. Following \cite{lin2024clickprompt}, we binarize the ratings with a threshold of 4 and use the 5-core setting, \ie all users and items have at least 5 interactions.

    \item \textbf{MovieLens-1M Dataset}\footnote{\url{https://grouplens.org/datasets/MovieLens/1m/}} is a benchmark movie recommendation dataset from Movielens with ratings ranging from 1 to 5. We binarize the ratings with a threshold of 4, while removing neutral samples with ratings equal to 3 following \cite{li2023ctrl}.
\end{itemize}
 The dataset statistics are summarized in Table~\ref{tab:datasets}. Following \cite{li2023ctrl, lin2024clickprompt}, we organize the review behaviors in ascending order of timestamps to partition each dataset into training, validation, and testing sets with ratios of 8:1:1 after preprocessing.

% \textbf{Metrics.}
\subsubsection{Evaluation Metrics}
Following previous work \cite{lin2024clickprompt,song2019autoint}, we employ two widely-used metrics \textbf{LogLoss} (binary cross-entropy loss) and \textbf{AUC} (area under the ROC curve) to evaluate performance. The \textbf{RelaImpr} \cite{yan2014coupled} (relative AUC improvement) is also reported. Notably, slightly higher AUC or lower LogLoss (e.g., \textbf{0.1\%}) can be regarded as significant improvement in CTR prediction, as indicated by previous studies \cite{huang2019fibinet,song2019autoint}.

% We also refer to [] 
% An increase of \textbf{0.001} in AUC (↑) or a decrease in Logloss (↓) can be considered significant, as indicated by previous studies \cite{li2023ctrl, sun2024large, zhou2018deep}. \textbf{RelaImpr}  is employed to measure the relative AUC improvement compared to the backbone CTR model \cite{zhou2018deep, li2023ctrl} and is defined as follows:
% \begin{equation}
% RelaImpr=(\frac{AUC(measure\ model)-0.5}{AUC(base\ model)-0.5}-1)\times100\%.
% \end{equation}

% \textbf{Baselines.}
\subsubsection{Baselines}
% For comparisons, we selected KAR (RecSys'24) \cite{xi2024towards}, LLM-CF (CIKM'24) \cite{sun2024large}, CTRL  (TORS'23) \cite{li2023ctrl} and EASE (CIKM'24) \cite{qiu2024ease} as LLM-enhanced baselines. These baselines are closely related and representative LLM-enhanced CTR methods. For fair comparisons, unless otherwise specified, we consistently used the Llama3-8B \cite{dubey2024llama} as the LLM. We also include two representative PLM-enhanced methods --- CELA (arXiv'24) \cite{wang2024cela} and  ClickPrompt (WWW'24) \cite{lin2024clickprompt} for comparisons. 
% For PLM-enhanced CTR methods, we selected  for comparisons.
For comparisons, we selected KAR (RecSys'24) \cite{xi2024towards}, LLM-CF (CIKM'24) \cite{sun2024large}, CTRL (TORS'23)  \cite{li2023ctrl} and EASE (CIKM'24) \cite{qiu2024ease} as LLM-enhanced baselines. These baselines are closely related and representative LLM-enhanced CTR methods. 
% For fair comparisons, unless otherwise specified, we consistently used the Llama3-8B \cite{dubey2024llama} as the LLM. 
We also include two representative PLM-enhanced methods --- CELA (arXiv'24) \cite{wang2024cela} and  ClickPrompt (WWW'24) \cite{lin2024clickprompt} for comparisons. 
% We integrated these baseline methods into the following representative traditional CTR models: 1) Classic FM-based methods: FM (ICDM'10) \cite{rendle2010factorization} and  DeepFM (IJCAI'17) \cite{guo2017deepfm}; 2) field-based CTR methods: FwFM (WWW'18) \cite{pan2018field} and FmFM (WWW'21) \cite{sun2021fm2}; 3) The state-of-the-art CTR methods:  FinalMLP (AAAI'23) \cite{mao2023finalmlp} and WuKong (ICML'24) \cite{zhang2024wukong}. The readers may refer to Appendix A for more details about these methods.

We integrated these baseline methods into the following representative traditional CTR models: 1) \textbf{Classic FM-based methods}: FM (ICDM'10) \cite{rendle2010factorization} and  DeepFM (IJCAI'17) \cite{guo2017deepfm}; 2) \textbf{Field-based CTR methods}: FwFM (WWW'18) \cite{pan2018field} and FmFM (WWW'21) \cite{sun2021fm2}; 3) \textbf{The state-of-the-art CTR methods}:  FinalMLP (AAAI'23) \cite{mao2023finalmlp} and WuKong (ICML'24) \cite{zhang2024wukong}. The readers may refer to Appendix~\ref{sec:appendix-experiments} for more details about these methods.

% \textbf{Implementation Details.}
\subsubsection{Implementation Details}
% A grid search is utilized to find the optimal hyperparameters. 
% For all compared methods, we closely refer to the configurations provided in their respective publications to ensure their optimal performance. For LLaCTR, $\lambda_{kl}$ and $\lambda_{fm}$ are tuned in \{0.01, 0.05, 0.1, 0.3, 0.5, 0.7, 1\}, and the feature sampling number in fine-tuning is set at 1000. 
% More implementation details are provided in Appendix A.

We use the  Adam \citep{kingma2014adam} optimizer to train all the  CTR models. The learning rate (lr) is set as 0.001 and the weight decay (wd) is tuned in \{1e-2, 1e-3, 1e-4, 1e-5, 1e-6, 0\}. 
We fix the feature embedding size to 32 and batch size to 4096 for all the backbones and datasets. 
For all compared methods, we closely refer to the configurations provided in their respective publications to ensure their optimal performance.
For LLaCTR fine-tuning, we fix the batch size as 128, the learning rate of LLM as 1e-4 and rank of LoRA \cite{hu2021lora} as 8, respectively. During LLaCTR enhancement, the $\lambda_{kl}$ and $\lambda_{fm}$ are tuned in \{0.01, 0.05, 0.1, 0.3, 0.5, 0.7, 1\}. We set the feature sampling number as 1000 by default, and set temperature parameter $\tau$ of contrastive loss into 0.02 following previous work \cite{wang2023improving}.
All methods are implemented in PyTorch and run on 8 Nvidia A800 GPUs. More implementation details are provided in the Appendix~\ref{sec:appendix-experiments}.

% where the detailed optimal hyperparameters for each method on each dataset and backbone are reported.

\subsection{Experimental Results}
In this section, we focus on performance comparison, efficiency analysis, and ablation studies. Additional results, including time complexity analysis, hyperparameter sensitivity analysis and more details, are provided in Appendix~\ref{sec:appendix-experiments-supplementary}.

\subsubsection{Performance Comparison (RQ1)}
% The overall experimental results are presented in Table~\ref{tab:performance}.
Table~\ref{tab:performance} shows the performance comparison of the proposed LLaCTR against the baseline methods. We observe that:

\textbf{Comparing with traditional CTR models.} LLaCTR can be applied to various types of basic CTR models, and yield performance gains in most cases. The improvements brought by LLaCTR are impressive, achieving an average AUC relative improvement of 2.24\% over all basic CTR models across four datasets. These results demonstrate the effectiveness of LLaCTR by injecting field semantic knowledge.
% 3) AUC performance of our LLaCTR surpasses all CTR backbone models in most cases (91.7\%), these results clearly validate the effectiveness of LLaCTR in injecting high quality field knowledge into traditional CTR models for enhancement.

\textbf{Comparing with PLM/LLM-enhanced CTR models.} We observe that LLaCTR outperforms two representative PLM-enhanced CTR models (CELA and ClickPrompt) and four LLM-enhanced CTR models (KAR, LLM-CF, CTRL and EASE) in most cases (89\%), which validates the effectiveness of our field-level enhancement paradigm. This result is highly surprising, as we only utilize the lightweight field knowledge rather than the cumbersome instance-level (or user/item-level) knowledge utilized by recent work. The reasons may be  1) field knowledge is indeed important and helpful for CTR prediction; 2) the lightweight field knowledge has been effectively exploited, enhancing both feature representation and feature interactions.

% easily digested by the CTR models, while the cumbersome semantic knowledge encoded from descriptions of instances or user/items seems exhibits difficulty for CTR models to digest. 3) We explicitly leverage the field knowledge in enhancing feature representation and feature interactions, two most important components on CTR models.  

% \subsection{Efficiency Study} 
% \label{sec:efficiency-study}

\subsubsection{Efficiency Analysis (RQ2)}
\label{sec:efficiency-study}
In this section, we conduct an in-depth analysis on the training efficiency of different LLM-enhanced CTR methods. In terms of inference efficiency, since all these methods pre-store LLM knowledge and ultimately depend on CTR models for recommendations, the variation is much smaller than that observed during training (\cf Table~\ref{tab:inference_time} in Appendix~\ref{sec:appendix-Inference-Efficiency}).

% \textbf{Empirical study.} Empirically, we analysis the total training time of representative conventional CTR models and LLM-enhanced CTR models on Amazon Video Games and MovieLens-1M datasets with millions of interactions. As shown in Figure~\ref{fig:motivation_empirical}, existing LLM-enhanced CTR methods (e.g., KAR, LLM-CF, CTRL and EASE) require over 290 times (on average) more computation time than conventional CTR models. With the increase in the number of interactions, this phenomenon will further exacerbate. 
% In real-world industrial scenarios with interaction scales numbering in the billions, these LLM-enhanced CTR models will face more significant efficiency challenges.

% \textbf{Efficiency comparison.}
The AUC performance and the total training time of compared methods are shown in Figure~\ref{fig:expxx}. The performances of LLM-enhanced methods are reported on their best backbones. We observe: 1) Compared to three representative LLM-enhanced methods, LLaCTR can improve efficiency by 10-100x and achieve better performance. 2) With the increase of the sampling number in the SSFT module, both the training time and AUC of LLaCTR will increase. This indicates increasing the number of samples used for self-supervised learning can yield higher-quality field semantic knowledge, but require more training time. 3) Even without fine-tuning, LLaCTR still achieves decent results. This suggests our LLaCTR could be applied in resource-constrained scenarios, enhancing existing CTR methods with incurring limited additional time cost. Besides, we also study the efficiency bottleneck and the time complexity of these LLM-enhanced methods, the readers may refer to Appendix~\ref{sec:appendix-experiments-time-complexity} for more details.

% However, the amount of fine-tuning data required in LLaCTR ( instructions) is much less than the LLM-CF, and it typically only needs 1 to 3 epochs of fine-tuning to get optimal results (Appendix~\ref{sec:appendix-experiments-hyperparameters}). 

\subsubsection{Ablation Study (RQ3)} 
We further conduct the ablation study, where the Self-supervised Fine-tuning (SSFT) module, Feature Representation Enhancement (FRE) module or Feature Interaction Enhancement (FIE) module is removed respectively. The results are presented in Table~\ref{tab:ablation-main}. 
As can be seen, both the three components are important — removing SSFT, FRE or the FIE module would result in performance drops in most case. Delving deeper into the SSFT module, we observe that self-supervised field-feature fine-tuning on LLMs is indeed helpful. The field description encoded by the fine-tuned LLM has higher quality than that extracted from the pre-trained corpus. For FRE module, we observe that the developed KL-Divergence loss is also important. It can effectively inject semantic knowledge of fields into the feature representations. For FIE module, we find that for both explicit feature modeling and implicit feature modeling CTR methods, injecting field interaction scores can effectively guide feature interaction and bring AUC improvement.

\begin{table}[t]
\centering
\caption{Ablation study on DeepFM and WuKong. Results on the other backbones are shown in Appendix~\ref{sec:appendix-ablation}.}
\vspace{-0.3cm}
\label{tab:ablation-main}
\scalebox{1}{
\begin{tabular}{@{}cc|cccc@{}}
\Xhline{1.2pt}
\toprule
\multicolumn{2}{c|}{Method} & Gift & Games & Music & Movie \\ \midrule
\multicolumn{1}{c|}{\multirow{5}{*}{DeepFM}} & Base & 0.6713 & 0.6692 & 0.7582 & 0.8342 \\
\multicolumn{1}{c|}{} & w/o FT & 0.6759 & 0.6732 & 0.7683 & 0.8352 \\
\multicolumn{1}{c|}{} & w/o $\lambda_{kl}$ & 0.6764 & 0.6699 & 0.7582 & 0.8343 \\
\multicolumn{1}{c|}{} & w/o $\lambda_{fm}$ & {0.6787} & {0.6741} & {0.7685} & {0.8351} \\
\multicolumn{1}{c|}{} & LLaCTR & \textbf{0.6797} & \textbf{0.6744} & \textbf{0.7687} & \textbf{0.8354} \\ \midrule
\multicolumn{1}{c|}{\multirow{5}{*}{WuKong}} & Base & 0.6728 & 0.6767 & 0.7514 & 0.8323 \\
\multicolumn{1}{c|}{} & w/o FT & {0.6800} & 0.6777 & {0.7558} & 0.8322 \\
\multicolumn{1}{c|}{} & w/o $\lambda_{kl}$ & 0.6730 & {0.6784} & 0.7524 & 0.8332 \\
\multicolumn{1}{c|}{} & w/o $\lambda_{fm}$ & 0.6763 & 0.6783 & 0.7494 & {0.8347} \\
\multicolumn{1}{c|}{} & LLaCTR & \textbf{0.6802} & \textbf{0.6819} & \textbf{0.7564} & \textbf{0.8397} \\ \bottomrule
\Xhline{1.2pt}
\end{tabular}
}
\vspace{-0.3cm}
\end{table}

\section{Related Work}
\label{sec:related_work}
\subsection{Traditional CTR Prediction}
% \textbf{Traditional CTR Prediction.}
CTR prediction is a core functional  module in personalized online services, where the key idea is to capture feature interaction patterns that capture the combinational relationships among multiple features. Traditional methods employ various operations for feature interaction, including product-based operators \cite{cheng2016wide, wang2021dcn}, convolutional networks \cite{liu2015convolutional}, and attention mechanisms  \cite{huang2019fibinet}. Recently, field-aware CTR methods  \cite{pan2018field, sun2021fm2} have gained attention, though they primarily rely on field IDs as categorical indicators without fully leveraging the semantic richness of field descriptions.
% CTR Prediction serves as a core function module in personalized online services \cite{ zhu2021open, cheng2016wide, cheng2020adaptive}. The key idea of traditional CTR prediction models is to capture the feature interaction patterns, which indicates the combination relationships of multiple features. 
% The feature interaction layer is generally implemented by a combination of various operations.
% Product operators include classical bi-level models \cite{chang2010training, qu2016product} and higher-order \cite{wang2017deep, lian2018xdeepfm, wang2021dcn} models.
% Convolutional operators \cite{liu2015convolutional,li2019fi,liu2019feature} utilize CNN \cite{gu2018recent} and GCN \cite{zhang2019graph} to capture the implicit high order interaction.
% Due to the success of attention mechanism, attention-based methods \cite{chen2021dcap,huang2019fibinet,song2019autoint} also achieved great success. 
% In recent years, several field-aware CTR methods have been proposed \cite{pan2018field, sun2021fm2}, however, these methods relied solely on field IDs as categorical indicators without leveraging the rich semantic information inherent in field descriptions.

% Recent years also witnessed some field-aware traditional CTR methods \cite{zhao2021fint, liu2018field, zhang2019field, qi2021deep, chen2019flen, zhang2019fat}. However, these methods mainly utilized field IDs as categorical indicators \cite{juan2016field, sun2021fm2,pan2018field}, rather than using the rich semantic information inherent in field descriptions.

\subsection{LLMs as CTR predictors} 
% \textbf{LLMs as CTR predictors.} 
Initial efforts explored the pre-trained language models as CTR predictors by reformulating recommendation tasks as NLP problems \cite{geng2022recommendation, cui2022m6,liu2022ptab}. With the advent of LLMs in content comprehension and semantic reasoning \cite{achiam2023gpt, dubey2024llama}, LLMs have demonstrated strong potential in CTR prediction \cite{bao2023tallrec,chen2024hllm}. However, LLMs suffer from high inference latency and resource demands. While acceleration techniques exist (e.g., BAHE \cite{geng2024breaking} for parameter reduction, Rella \cite{lin2024rella} for sequence shortening), their computational overhead remains prohibitive for real-world deployment due to massive model sizes.

\subsection{LLM-enhanced CTR Prediction} 
% \textbf{LLM-enhanced CTR Prediction.} 
% Existing studies are focusing on leveraging the open-world knowledge, semantic encoding and text generation capabilities of LLMs to enhance traditional CTR models \cite{li2023ctrl, zhang2023collm, xi2024towards, sun2024large, salemi2024optimization}. 
% Some work leveraged pretrained language model (like BERT \cite{devlin2019bert}) to model feature interactions \cite{muhamed2022dcaf, tian2023ufin, muhamed2021ctr, wang2023bert4ctr, liu2022ptab, lin2024clickprompt, wang2024cela}, such as P-tab \cite{liu2022ptab}, ClickPrompt \cite{lin2024clickprompt}, CELA\cite{wang2024cela} and so on. 
% For example, ClickPrompt \cite{lin2024clickprompt} uses a CTR model to encode instances into soft tokens, which are then trained as prefix tokens for a PLM for generating instance-level semantic knowledge.
To address the efficiency challenges inherent in LLM-based CTR predictors, the LLM-enhanced CTR paradigm has been extensively investigated as an alternative approach. This paradigm primarily focuses on leveraging the powerful semantic encoding and knowledge reasoning capabilities of LLMs to enhance traditional CTR models \cite{li2023ctrl,qiu2024ease}. For example, KAR \cite{xi2024towards} leverages LLMs to enhance the knowledge of users and items. LLM-CF \cite{sun2024large} uses LLMs to generate chains of thought (CoT) for enhancement. CTRL \cite{li2023ctrl} utilizes the LLM as the feature encoder for instances and uses contrastive learning to align the knowledge. EASE \cite{qiu2024ease} trains a semantic adaptor to align item embeddings with LLM. Other work also focuses on specific CTR scenarios, such as cross-domain\cite{fu2023unified} and  interpretability \cite{yu2024explainable}.
Despite their promising performance, they involve enhancements at the instance-level or user/item-level, which results in great inefficiency issue in dealing with large-scale instances or entities in practice. Our approach is applied on the more efficient and cost-effective field-level, without incurring superabundant computational overhead.

\subsection{Other Related Work} 
There are also two related topics: 1) \textbf{PLM-enhanced CTR methods}: Early studies have leveraged pretrained language models (PLMs, such as BERT \cite{devlin2019bert}) to enhance CTR models, e.g., CELA \cite{wang2024cela} pretrains PLMs on item features and aligns them with ID embeddings for enhancement. ClickPrompt \cite{lin2024clickprompt} uses the ID embeddings as prefix soft tokens in PLMs for generating instance-level knowledge. However, the capacity of early PLMs is generally inferior to that of recent LLMs, and recent studies have demonstrated the superior performance of LLM-enhanced methods. Consequently, our work aims to further investigate and advance the LLM-enhanced CTR modeling. 2) \textbf{LLM-enhanced recommendation}: Recent years have also witnessed the integration of LLMs on other recommendation tasks \cite{cui2025hatllm,wang2025msl,wang2026doesllmfocusright}, \eg sequential recommendation \cite{cui2024distillation,wang2025llm4dsr}, collaborative filtering \cite{ren2024representation,lin2025recommendation}. Notably, CTR prediction differs from these tasks in that it typically involves various features and emphasizes modeling feature interactions. Thus, our LLaCTR is specifically designed to improve feature representation and interactions in CTR prediction.

\section{Conclusion}
This work proposes a lightweight method, LLaCTR, which leverages LLMs to extract semantic knowledge from feature fields for enhancing CTR models. Specifically, LLaCTR employs a self-supervised field-feature fine-tuning to distill high-quality field-level knowledge, which is subsequently utilized to improve both feature representation and feature interaction modeling. Extensive experiments demonstrate the superiority of LLaCTR in both predictive accuracy and computational efficiency. In the future, it would be valuable to investigate more sophisticated strategies for integrating field semantics to further enhance CTR methods.

\begin{acks}
This work is supported by the Zhejiang Province “JianBingLingYan+X” Research and Development Plan (2025C02020). We thank the reviewers for their valuable and insightful suggestions that improve the paper.
\end{acks}

\newpage
%%
%% The next two lines define the bibliography style to be used, and
%% the bibliography file.
\bibliographystyle{ACM-Reference-Format}
\bibliography{sample-base}

%%
%% If your work has an appendix, this is the place to put it.
\appendix
% Experimental Details --------------------------------------
\section{Appendix A: Experimental Details} 
\label{sec:appendix-experiments}

\subsection{Baselines} 
\label{sec:appendix-experiments-baselines}

We reproduced the following LLM-enhanced CTR methods as baselines in our experiments:

\begin{itemize}[topsep=0pt,leftmargin=10pt]
    \setlength{\itemsep}{0pt}
    \item \textbf{KAR} (RecSys'24) \citep{xi2024towards}: KAR leverages the LLM as a open-word knowledge base to enhance the profile of users and items, and then pass the encoded user/item embeddings through a knowledge adaptation network to enhance traditional CTR models.   
    \item \textbf{LLM-CF} (CIKM'24) \citep{sun2024large}: LLM-CF uses LLMs to generate a base of chains of thought (CoT) for sampled instances, which are further retrieved to enhance CTR models. Consistent with the original paper, we sampled 10\% of the instances as sampled data for fine-tuning and CoT data generation. 
    \item \textbf{CTRL} (TORS'23) \citep{li2023ctrl}: CTRL utilizes the LLM as the feature encoder for instances and uses a fine-grained contrastive learning scheme to align cross-modal knowledge. 
    \item \textbf{EASE} (CIKM'24) 
    \citep{qiu2024ease}: EASE trains a semantic adaptor network to align item embeddings with the LLM, and then use it for encoding during inference. Following the original paper, we inited the semantic adapter using BERT-Base \cite{devlin2019bert} with 12 transformer blocks, and cross-attention layers are added to the 6-th and 12-th transformer blocks.
\end{itemize}
% , and during downstream CTR prediction, we also retrieve the Top 4 samples closest in semantics from the CoT dataset as enhanced knowledge.
% In our implementation, we used Llama3-8B \cite{dubey2024llama} as the LLM to encode instances embeddings for fair comparison.

For fair comparisons, unless otherwise specified, we consistently used the Llama3-8B \cite{dubey2024llama} as the LLM in all the LLM-enhanced CTR methods. Besides, we also reproduced two representative PLM-enhanced methods:
\begin{itemize}[topsep=0pt,leftmargin=10pt]
    \item \textbf{CELA} (arXiv'24) 
    \citep{wang2024cela}: CELA first pretrains a PLM on item descriptions, then aligns item textual embeddings with pretrained ID-based embeddings, and finally merges them into CTR models for enhancement. We selected Robert-base \cite{liu2019roberta} as the PLM as it performs the best in the original paper. 
    \item \textbf{ClickPrompt} (WWW'24) \citep{lin2024clickprompt}: ClickPrompt uses a CTR model to encode instances into soft tokens, which are then treated as prefix tokens in a PLM for generating instance-level semantic knowledge. Following the original paper, we selected Robert-base \cite{liu2019roberta} as the PLM in our experiment.
\end{itemize}
% Backbones ------------------------------------------------
\subsection{Backbones} \label{sec:appendix-experiments-backbones}
For all the backbone models, we reused the implement of an open-source CTR prediction library\footnote{\url{https://github.com/reczoo/FuxiCTR}} - FuxiCTR \cite{zhu2022bars}, following the previous work \cite{mao2023finalmlp,zhu2023final,li2024dcnv3}.
We selected six representative CTR methods as backbones  in our experiments, including:

\begin{itemize}[topsep=0pt,leftmargin=10pt]
    \setlength{\itemsep}{0pt}
    \item \textbf{FM (ICDM'10)} \cite{rendle2010factorization}: FM introduces factorization machines to model pairwise feature interactions via factorized parameters. 
    \item \textbf{DeepFM (IJCAI'17)} \cite{guo2017deepfm}: DeepFM combines factorization machines with deep neural networks to jointly learn low-order and high-order feature interactions. We fix the DNN layers at [300, 300, 128] in our experiments.
    \item \textbf{FwFM (WWW'18)} \cite{pan2018field}: FwFM proposes field-weighted factorization machines that learn field-specific interaction weights to better capture heterogeneous feature relationships.
    \item \textbf{FmFM (WWW'21)} \cite{sun2021fm2}: FmFM enhances FM with field-matrix factorization using learnable projection matrices per field pair, enabling more expressive cross-field interactions.
    \item \textbf{FinalMLP (AAAI'23)} \cite{mao2023finalmlp}: FinalMLP designs two Stream Feature Interaction networks to progressively refine feature representations learned by MLP. We select the MLP hidden units size from \{400, 500\} and the MLP layer number from \{2, 3\}.
    \item \textbf{WuKong (ICML'24)} \cite{zhang2024wukong}: WuKong employs an interaction stack on the WuKong layer to capture feature interactions, where each WuKong layer consists of a Factorization Machine Block (FMB) and a Linear Compress Block. We select the interaction layer number from \{4, 8\}, the FMB layer from \{2, 3\} and the compression dimension from \{24, 32\} respectively.
\end{itemize}
% the FMB size from \{128, 200\},
For the implementation of LLaCTR, the SSFT module and FRE module remained consistent across all backbones. The implementation of the FIE module had slightly difference on explicit and implicit feature interaction backbones. Specifically:
\begin{itemize}[topsep=0pt,leftmargin=10pt]
    \setlength{\itemsep}{0pt}
    \item 1) For explicit feature interaction backbones (FM, DeepFM, FwFM and FmFM), the FIE module was implemented by adding the learnable field interaction matrix on the original feature interaction matrix. By explicitly guiding the learning of feature interactions modeling, the FIE module can directly inject the field interaction knowledge into the CTR models. 
    \item 2) For implicit feature interaction backbones (FinalMLP and WuKong), the FIE module was treated as an additional plugin network to explicitly influence the second-order feature interactions learning and generate CTR prediction logit, which was later fused into the final prediction result of the original CTR model.
\end{itemize}

\begin{table}[t]
\centering
\caption{Fine-tuning loss  (in SSFT module) comparison between language generative loss ($\mathcal{L}_{LG}$) and contrastive loss  ($\mathcal{L}_{CL}$).}
\vspace{-0.3cm}
\label{tab:appendix-ft-loss}
\scalebox{0.78}{
\begin{tabular}{cc|ccc|ccc}
\Xhline{1.2pt}
\hline
\multicolumn{2}{c|}{\multirow{2}{*}{Method}} & \multicolumn{3}{c|}{Gift Cards} & \multicolumn{3}{c}{Video Games} \bigstrut\\ \cline{3-8} 
\multicolumn{2}{c|}{} & Logloss ↓ & AUC ↑ & RelaImpr & Logloss ↓ & AUC ↑ & RelaImpr \bigstrut \\ \hline
\multicolumn{1}{c|}{\multirow{3}{*}{DeepFM}} & Base & 0.4751 & 0.6713 &  & 0.5762 & 0.6692 &  \bigstrut[t]\\
\multicolumn{1}{c|}{} & $\mathcal{L}_{LG}$ & 0.4659 & 0.6794 & 4.73\% & \textbf{0.5755} & 0.6718 & 1.55\% \\
\multicolumn{1}{c|}{} & $\mathcal{L}_{CL}$ & \textbf{0.4657} & \textbf{0.6797} & \textbf{4.90\%} & 0.5834 & \textbf{0.6744} & \textbf{3.10\%} \bigstrut[b]\\ \hline
\multicolumn{1}{c|}{\multirow{3}{*}{WuKong}} & Base & 0.4755 & 0.6728 &  & 0.5819 & 0.6767 &  \bigstrut[t]\\
\multicolumn{1}{c|}{} & $\mathcal{L}_{LG}$ & 0.4748 & 0.6800 & 4.14\% & 0.5795 & 0.6795 & 1.57\% \\
\multicolumn{1}{c|}{} & $\mathcal{L}_{CL}$ & \textbf{0.4745} & \textbf{0.6802} & \textbf{4.24\%} & \textbf{0.5763} & \textbf{0.6819} & \textbf{2.91\%} \bigstrut[b]\\ \hline
\Xhline{1.2pt}
\end{tabular}
}
\vspace{-0.3cm}
\end{table}

\begin{table}[t]
\centering
\caption{Feature representation enhancement loss  (in FRE module) comparison between mean squared error loss ($\mathcal{L}_{MSE}$), contrastive loss  ($\mathcal{L}_{CL}$), and  Kullback-Leibler divergence loss ($\mathcal{L}_{KL}$).}
\vspace{-0.3cm}
\label{tab:appendix-fre-loss}
\scalebox{0.78}{
\begin{tabular}{cc|ccc|ccc}
\Xhline{1.2pt}
\hline
\multicolumn{2}{c|}{\multirow{2}{*}{Method}} & \multicolumn{3}{c|}{Gift Cards} & \multicolumn{3}{c}{Video Games} \\ \cline{3-8}
\multicolumn{2}{c|}{} & Logloss ↓ & AUC ↑ & RelaImpr & Logloss ↓ & AUC ↑ & RelaImpr \bigstrut \\  \hline
\multicolumn{1}{c|}{\multirow{4}{*}{DeepFM}} & Base & 0.4751 & 0.6713 &  & 0.5762 & 0.6692 &  \bigstrut[t]\\
\multicolumn{1}{c|}{} & $\mathcal{L}_{MSE}$ & 0.4750 & 0.6767 & 3.15\% & \textbf{0.5760} & 0.6711 & 1.15\% \\
\multicolumn{1}{c|}{} & $\mathcal{L}_{CL}$ & 0.4750 & 0.6763 & 2.92\% & 0.5762 & 0.6699 & 0.44\% \\
\multicolumn{1}{c|}{} & $\mathcal{L}_{KL}$ & \textbf{0.4657} & \textbf{0.6797} & \textbf{4.90\%} & 0.5834 & \textbf{0.6744} & \textbf{3.10\%} \bigstrut[b]\\ \hline
\multicolumn{1}{c|}{\multirow{4}{*}{WuKong}} & Base & 0.4755 & 0.6728 &  & 0.5819 & 0.6767 &  \bigstrut[t]\\
\multicolumn{1}{c|}{} & $\mathcal{L}_{MSE}$ & 0.4755 & 0.6737 & 0.49\% & 0.5787 & 0.6795 & 1.57\% \\
\multicolumn{1}{c|}{} & $\mathcal{L}_{CL}$ & 0.4753 & 0.6738 & 0.55\% & 0.5805 & 0.6788 & 1.17\% \\
\multicolumn{1}{c|}{} & $\mathcal{L}_{KL}$ & \textbf{0.4745} & \textbf{0.6802} & \textbf{4.24\%} & \textbf{0.5763} & \textbf{0.6819} & \textbf{2.91\%} \bigstrut[b]\\ \hline
\Xhline{1.2pt}
\end{tabular}
}
\vspace{-0.3cm}
\end{table}

%%%%%%%%%%%%%%%%%%%%%%%%%%%%%%%%%%%%%%%%%%%%%%%%%%%%%%%%%%%%
% Supplementary experiments --------------------------------
\section{Appendix B: Supplementary Experiments} 
\label{sec:appendix-experiments-supplementary}

\subsection{Efficiency Study}
\label{sec:appendix-experiments-time-complexity}
\subsubsection{Efficiency Bottleneck Analysis}
The training time of all LLM-enhanced CTR methods can be divided into four parts: \textit{fine-tuning}, \textit{knowledge generation}, \textit{semantic encoding} and \textit{CTR prediction}. To study the efficiency bottleneck, we conducted a detailed training time analysis on  Video Games and MovieLens-1M datasets. The results are shown in Table~\ref{tab:Running-Time-Details}. We can derive:
\begin{itemize}[topsep=0pt,leftmargin=10pt]
\setlength{\itemsep}{0pt}

\item 1) The efficiency bottleneck of KAR and LLM-CF is the knowledge generation. KAR requires to generate knowledge for a large number of users/items, while LLM-CF needs to generate the CoT data on numerous instances (even if sampled as 10\%). The reasoning and generation of LLMs consume a significant amount of time, and rise constantly as the length of generated text increases. 

\item 2) The encoding time overhead of CTRL is significantly higher than other methods. Although encoding usually costs less time than Fine-tuning and Inference, CTRL still faces an efficiency bottleneck due to the large number of training instances that need to be encoded.

\item 3) The time overhead of fine-tuning with EASE is higher than other methods. EASE freezes the LLM and trains trains the semantic adaptor network during fine-tuning, but the frozen LLM still needs to be involved in the model forward phase. Besides, the parameter size of EASE's semantic adaptor network (a BERT-Base \cite{devlin2019bert} network with 12 transformer blocks and an additional 2 cross-attention layers) is about 110M, while the trainable parameter size of LoRA \cite{hu2021lora} based Llama3-8B \cite{dubey2024llama} (used in LLM-CF and LLaCTR) is only approximately 3.4M (with the rank LoRA as 8), leading to a higher fine-tuning time overhead for EASE.

\item 4) The efficiency bottleneck of LLaCTR lies in fine-tuning. Although fine-tuning on the sampled field-feature data (in the SSFT module of LLaCTR) still incurs considerable time overhead compared to traditional CTR methods, it is much more efficient than fine-tuning on the numerous instance level or user/item level data (e.g., LLM-CF and EASE). 
\end{itemize}

\begin{table}[t]
\centering
\caption{Time Cost (s) details of LLM-enhanced CTR baselines and our LLaCTR.}
\vspace{-0.3cm}
\label{tab:Running-Time-Details}
\scalebox{0.8}{
\begin{tabular}{c|c|cccc}
\Xhline{1.2pt}
\hline
\multirow{2}{*}{Dataset} & \multirow{2}{*}{ Method} & \multicolumn{4}{c}{   Time Cost in Details}  \bigstrut\\ \cline{3-6}
 &  & \multicolumn{1}{c|}{Fine-tuning} & \multicolumn{1}{c|}{Generation} & \multicolumn{1}{c|}{Enoding} & Prediction   \bigstrut\\ \hline
\multirow{5}{*}{Video   Games} & KAR & \multicolumn{1}{c|}{0} & \multicolumn{1}{c|}{77,861} & \multicolumn{1}{c|}{3,886} & 120  \bigstrut[t]\\
 & LLM-CF & \multicolumn{1}{c|}{15,179} & \multicolumn{1}{c|}{78,104} & \multicolumn{1}{c|}{1,676} & 178 \\
 & CTRL & \multicolumn{1}{c|}{0} & \multicolumn{1}{c|}{0} & \multicolumn{1}{c|}{5,839} & 329 \\
 & EASE & \multicolumn{1}{c|}{32,221} & \multicolumn{1}{c|}{0} & \multicolumn{1}{c|}{1,747} & 125 \\
 & LLaCTR & \multicolumn{1}{c|}{599} & \multicolumn{1}{c|}{0} & \multicolumn{1}{c|}{12} & 124  \bigstrut[b]\\ \hline
\multirow{5}{*}{MovieLens-1M} & KAR & \multicolumn{1}{c|}{0} & \multicolumn{1}{c|}{2,057} & \multicolumn{1}{c|}{108} & 42  \bigstrut[t]\\
 & LLM-CF & \multicolumn{1}{c|}{5,234} & \multicolumn{1}{c|}{26,094} & \multicolumn{1}{c|}{355} & 64  \\
 & CTRL & \multicolumn{1}{c|}{0} & \multicolumn{1}{c|}{0} & \multicolumn{1}{c|}{1,401} & 141  \\
 & EASE & \multicolumn{1}{c|}{12,813} & \multicolumn{1}{c|}{0} & \multicolumn{1}{c|}{49} & 43  \\
 & LLaCTR & \multicolumn{1}{c|}{415} & \multicolumn{1}{c|}{0} & \multicolumn{1}{c|}{11} & 34  \bigstrut[b]\\ \hline
 \Xhline{1.2pt}
\end{tabular}
}
\vspace{-0.5cm}
\end{table}

\subsubsection{Time Complexity Analysis} 
To further study the training efficiency of LLaCTR and other LLM-enhanced CTR methods, we discuss the training time complexity of these models here. 
% Let $A$ represent the average input token sequence length of the LLM, $B$ represent the average output token sequence length of the LLM, and $D$ represent the embedding dimension of the LLM. The average time overhead of the LLM during fine-tuning and encoding can be expressed as $H(A)$, while the average time overhead during generation can be expressed as $G(A, B)$. For the original CTR model, let $t$ represent the average time overhead of the model and $d$ represent the feature embedding dimension. We use $U$ to represent the number of users, $I$ to represent the number of items, $N$ to represent the total number of instances, respectively. 
For the LLMs, let $A$ and $B$ denote the average input and output token sequence lengths of the LLM, respectively, and $D$ its embedding dimension. The average time overhead of the LLM can be expressed as $H(A)$ for fine-tuning/encoding and $G(A,B)$ for generation. For the original CTR model, $t$ denotes average time overhead and $d$ the feature embedding dimension. $U$, $I$, and $N$ denote the number of users, items, and total instances, respectively. 

In LLaCTR, let $K$ represent the number of fields and $S$ represent the feature sample number of SSFT module. The time complexity of the SSFT module is $O(SKH(A) + KH(A))$, while the FRE module is $O(KDd)$ and the FRE module is $O(t + K^2)$, respectively. The overall time complexity of LLaCTR is $O((S + 1)KH(A) + KDd + (t + K^2))$. Since the time overhead on the LLM is much greater than the enhancement on CTR models, the SSFT module will dominate the overall time overhead, and the time complexity can be simplified to $O((S + 1)KH(A))$.

Similarly, the time complexity of KAR is $O((U + I)G(A, B) + (U + I)H(A))$, LLM-CF is $O(2NH(A) + NG(A, B))$, CTRL is $O(NH(A))$, and EASE is $O(2IH(A))$. Given that $K < SK \ll (U+I) < N$, LLaCTR exhibits significantly lower training time complexity compared to other LLM-enhanced CTR methods. This further highlights the efficiency of LLaCTR’s field-level enhancement approach.

\subsection{Empirical Study}

\subsubsection{Fine-tuning Loss Study}
\label{sec:appendix-Fine-tuning}
For the fine-tuning loss in the SSFT module, we compare the performance of language generative loss ($\mathcal{L}_{LG}$) with contrastive loss  ($\mathcal{L}_{CL}$) on over DeepFM and WuKong backbones, and the results are shown in Table~\ref{tab:appendix-ft-loss}. We observe that the contrastive loss performs slightly better than the language generative loss, so we use the contrastive loss as the final fine-tuning loss.

\begin{table}[t]
\centering
\caption{Ablation Study on the other backbones.}
\vspace{-0.3cm}
\label{tab:ablation}
\scalebox{0.9}{
\begin{tabular}{cc|cccc}
\Xhline{1.2pt}
\hline
\multicolumn{2}{c|}{\multirow{1}{*}{Method}} & \multicolumn{1}{c}{Gift} & \multicolumn{1}{c}{Games} & \multicolumn{1}{c}{Music} & \multicolumn{1}{c}{Movie} \bigstrut\\ \hline
% \multicolumn{2}{c|}{Method} & Gift & Games & Music & Movie \\ \hline
\multicolumn{1}{c|}{\multirow{5}{*}{FwFM}} & Base & 0.6691 & 0.6683 & 0.7553 & 0.8365 \bigstrut[t]\\
\multicolumn{1}{c|}{} & w/o FT & {0.6714} & {0.6757} & \textbf{0.7587} & 0.8365 \\
\multicolumn{1}{c|}{} & w/o FRE & 0.6692 & 0.6729 & 0.7553 & {0.8366} \\
\multicolumn{1}{c|}{} & w/o FIE & 0.6710 & 0.6652 & 0.7572 & 0.8364 \\
\multicolumn{1}{c|}{} & LLaCTR & \textbf{0.6718} & \textbf{0.6780} & {0.7585} & \textbf{0.8367} \bigstrut[b]\\ \hline
\multicolumn{1}{c|}{\multirow{5}{*}{FmFM}} & Base & 0.6747 & 0.6706 & 0.7628 & 0.8325 \bigstrut[t]\\
\multicolumn{1}{c|}{} & w/o FT & {0.6759} & {0.6759} & {0.7614}  & 0.8334 \\
\multicolumn{1}{c|}{} & w/o FRE & 0.6750 & 0.6714 & 0.7608 & {0.8334} \\
\multicolumn{1}{c|}{} & w/o FIE & 0.6753 & 0.6710 & 0.7612 & 0.8332 \\
\multicolumn{1}{c|}{} & LLaCTR & \textbf{0.6761} & \textbf{0.6765} & \textbf{0.7615} & \textbf{0.8335} \bigstrut[b]\\ \hline
\multicolumn{1}{c|}{\multirow{5}{*}{FinalMLP}} & Base & 0.6723 & 0.6761 & 0.7557 & 0.8260 \bigstrut[t]\\
\multicolumn{1}{c|}{} & w/o FT & {0.6815} & {0.6787} & 0.7554 & 0.8330 \\
\multicolumn{1}{c|}{} & w/o FRE & 0.6757 & 0.6772 & \textbf{0.7564} & 0.8332 \\
\multicolumn{1}{c|}{} & w/o FIE & 0.6800 & 0.6775 & {0.7555} & 0.8324 \\
\multicolumn{1}{c|}{} & LLaCTR & \textbf{0.6818} & \textbf{0.6832} & 0.7552 & \textbf{0.8340} \bigstrut[b]\\ \hline
\Xhline{1.2pt}
\end{tabular}
}
\vspace{-0.3cm}
\end{table}

\subsubsection{Alignment Loss Study}
\label{sec:appendix-Alignment}
For the feature representation enhancement loss in the FRE module, we compare the performance of mean squared error loss ($\mathcal{L}_{MSE}$), contrastive loss  ($\mathcal{L}_{CL}$), and  Kullback-Leibler divergence loss ($\mathcal{L}_{KL}$) over DeepFM and WuKong backbones, and the results are shown in Table~\ref{tab:appendix-fre-loss}. We find that the KL divergence loss performs better than the MSE loss and CL loss, which is more suitable for enhancement loss.

\subsection{Inference Efficiency Study}
\label{sec:appendix-Inference-Efficiency}

\begin{table}[t]
\centering
\caption{Inference Time (s) on LLM-enhanced CTR methods.}
\vspace{-0.3cm}
\label{tab:inference_time}
\renewcommand{\arraystretch}{1.2}
% \scriptsize
\scalebox{0.85}{
\begin{tabular}{c|cccccc}
\Xhline{1.2pt}
Dataset & WuKong & KAR & LLM-CF & CTRL & EASE & LLaCTR \\ 
\hline
Video Games  & 1.57 & 1.66 & 1.84 & 1.81 & 1.73 & 1.63 \\
MovieLens-1M  & 0.32 & 0.41 & 0.53 & 0.46 & 0.45 & 0.41 \\
\Xhline{1.2pt}
\end{tabular}
}
\vspace{-0.3cm}
\end{table}

We compare the total inference time of different LLM-enhanced methods on the Video Games and MovieLens-1M datasets, with the results shown in Table~\ref{tab:inference_time} (WuKong serves as the backbone model). It can be observed that the differences among various LLM-enhanced methods during the inference stage are negligible and do not introduce significant latency compared to the backbone model. This is because these LLM-enhanced methods have pre-stored the knowledge of LLMs, requiring only lightweight networks and a small number of parameters to inject the knowledge into the inference process of traditional CTR models, thereby avoiding excessive additional latency.

\subsection{Ablation Study}
\label{sec:appendix-ablation}
The ablation study results on the FwFM, FmFM and FinalMLP backbones are presented in Table~\ref{tab:ablation}. As can be seen, the results are consistent with those obtained from other backbones. Both the three modules are important, removing each would result in performance drops.

\subsection{Hyperparameters Sensitivity} 

\begin{figure}[ht]
    % \centering
    % \begin{subfigure}[b]{0.235\textwidth}
    % \captionsetup{labelformat=empty}
    %     \includegraphics[width=\textwidth]{fig/lam_kl_Gift_Cards_DeepFM.pdf}
    % \vspace{-0.1cm}
    % \caption{$\quad\quad \lambda_{kl}$}
    % \end{subfigure}
    % \begin{subfigure}[b]{0.235\textwidth}
    %     \includegraphics[width=\textwidth]{fig/lam_kl_Gift_Cards_FmFM.pdf}
    %     \caption{$\lambda_{kl}$}
    % \end{subfigure}
    % \begin{subfigure}[b]{0.235\textwidth}
    %     \includegraphics[width=\textwidth]{fig/lam_kl_Gift_Cards_WuKong.pdf}
    %     \caption{$\lambda_{kl}$}
    % \end{subfigure}    
    % \begin{subfigure}[b]{0.235\textwidth}
    %     \includegraphics[width=\textwidth]{fig/lam_fm_Gift_Cards_DeepFM.pdf}
    %     \caption{$\lambda_{fm}$}
    % \end{subfigure}
    % \begin{subfigure}[b]{0.235\textwidth}
    %     \includegraphics[width=\textwidth]{fig/lam_fm_Gift_Cards_FmFM.pdf}
    %     \caption{$\lambda_{fm}$}
    % \end{subfigure}
    % \begin{subfigure}[b]{0.235\textwidth}
    %     \includegraphics[width=\textwidth]{fig/lam_fm_Gift_Cards_WuKong.pdf}
    %     \caption{$\lambda_{fm}$}
    % \end{subfigure}
    
    \centering
    \begin{subfigure}[b]{0.235\textwidth}
    \captionsetup{labelformat=empty}
        \includegraphics[width=\textwidth]{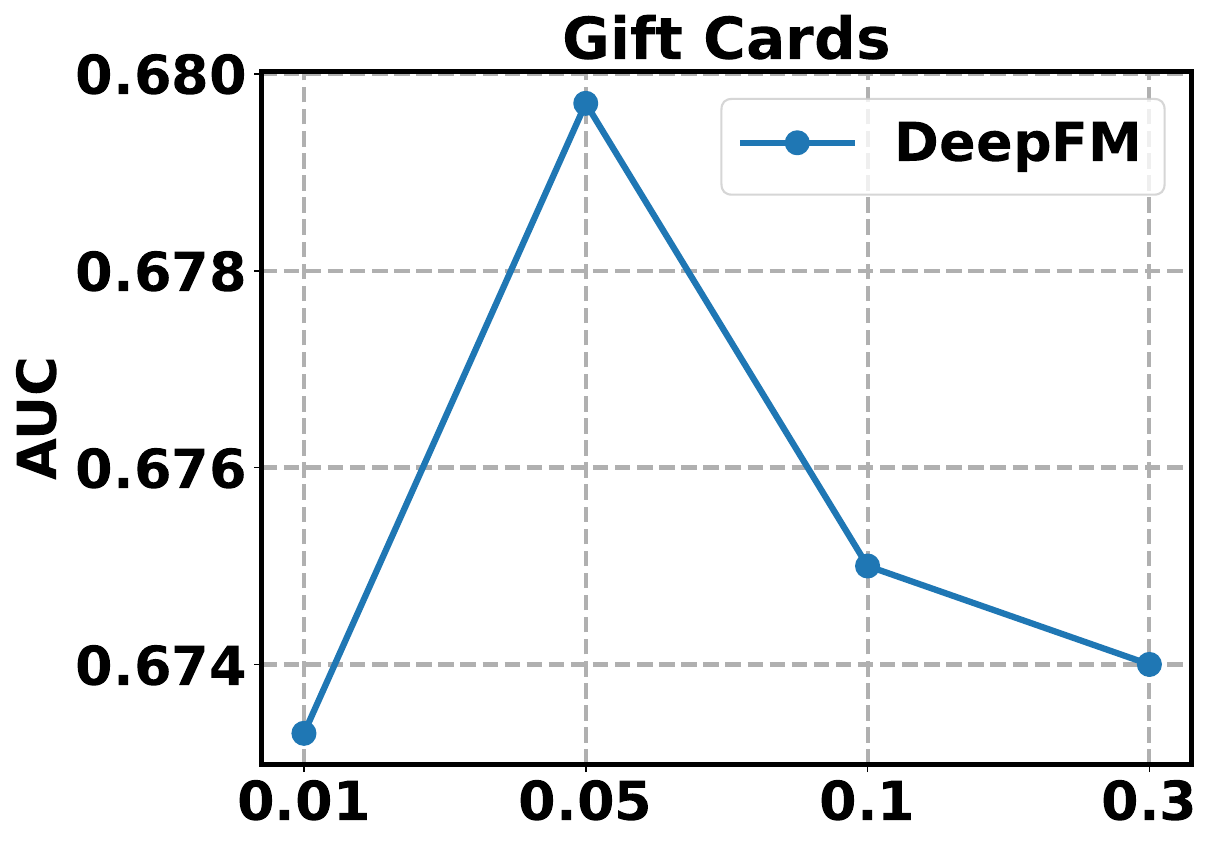}
        % \vspace{-0.1cm}
        \caption{$\quad\quad \lambda_{kl}$}
    \end{subfigure}
    \begin{subfigure}[b]{0.235\textwidth}
    \captionsetup{labelformat=empty}
        \includegraphics[width=\textwidth]{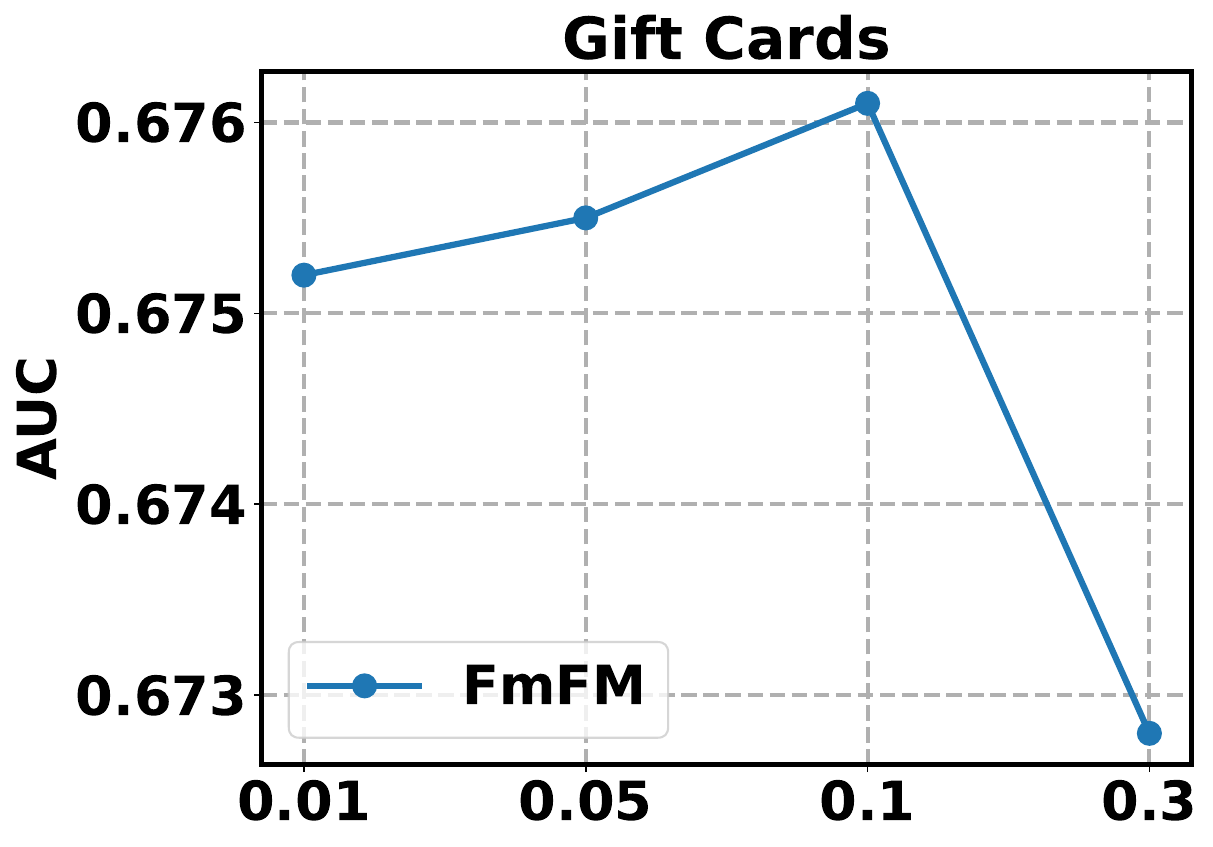}
        % \vspace{-0.1cm}
        \caption{$\quad\quad \lambda_{kl}$}
    \end{subfigure}
    \begin{subfigure}[b]{0.235\textwidth}
    \captionsetup{labelformat=empty}
        \includegraphics[width=\textwidth]{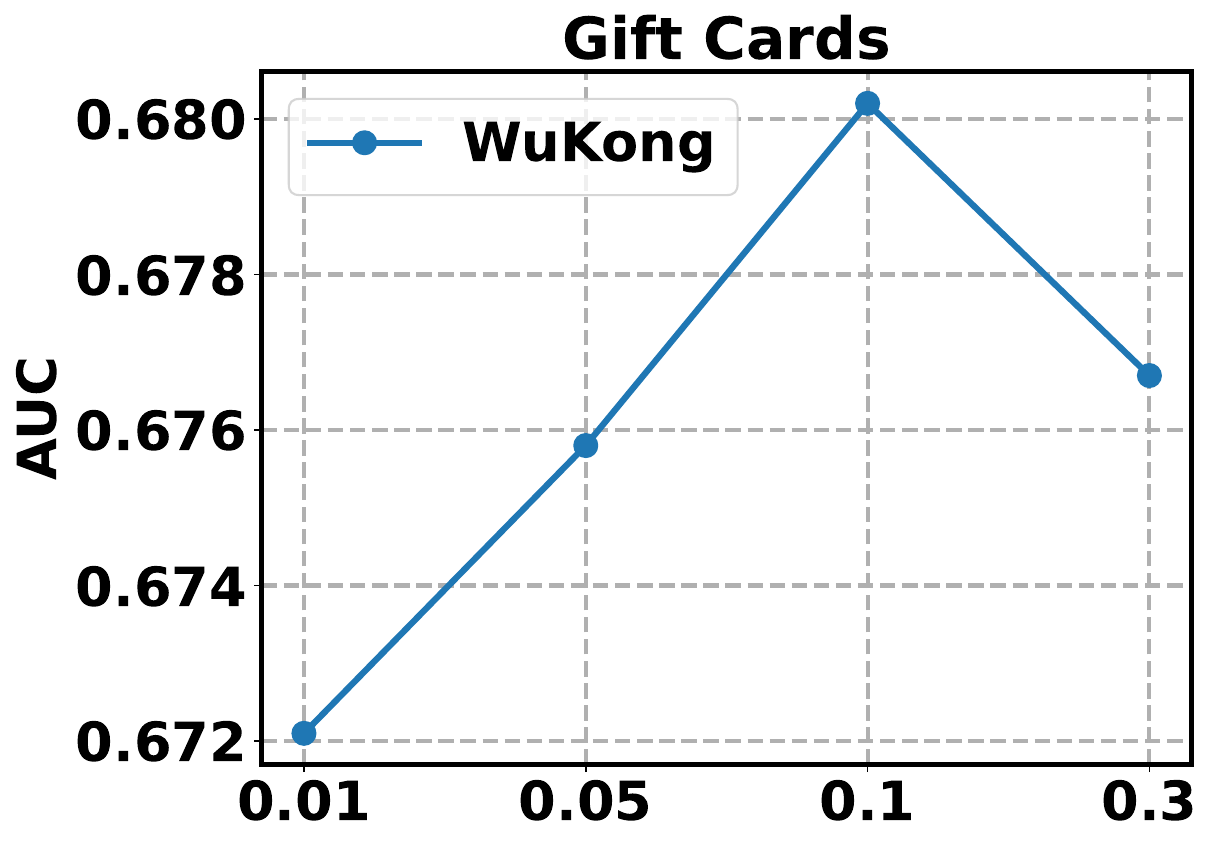}
        \caption{$\quad\quad \lambda_{kl}$}
    \end{subfigure}    
    \begin{subfigure}[b]{0.235\textwidth}
    \captionsetup{labelformat=empty}
        \includegraphics[width=\textwidth]{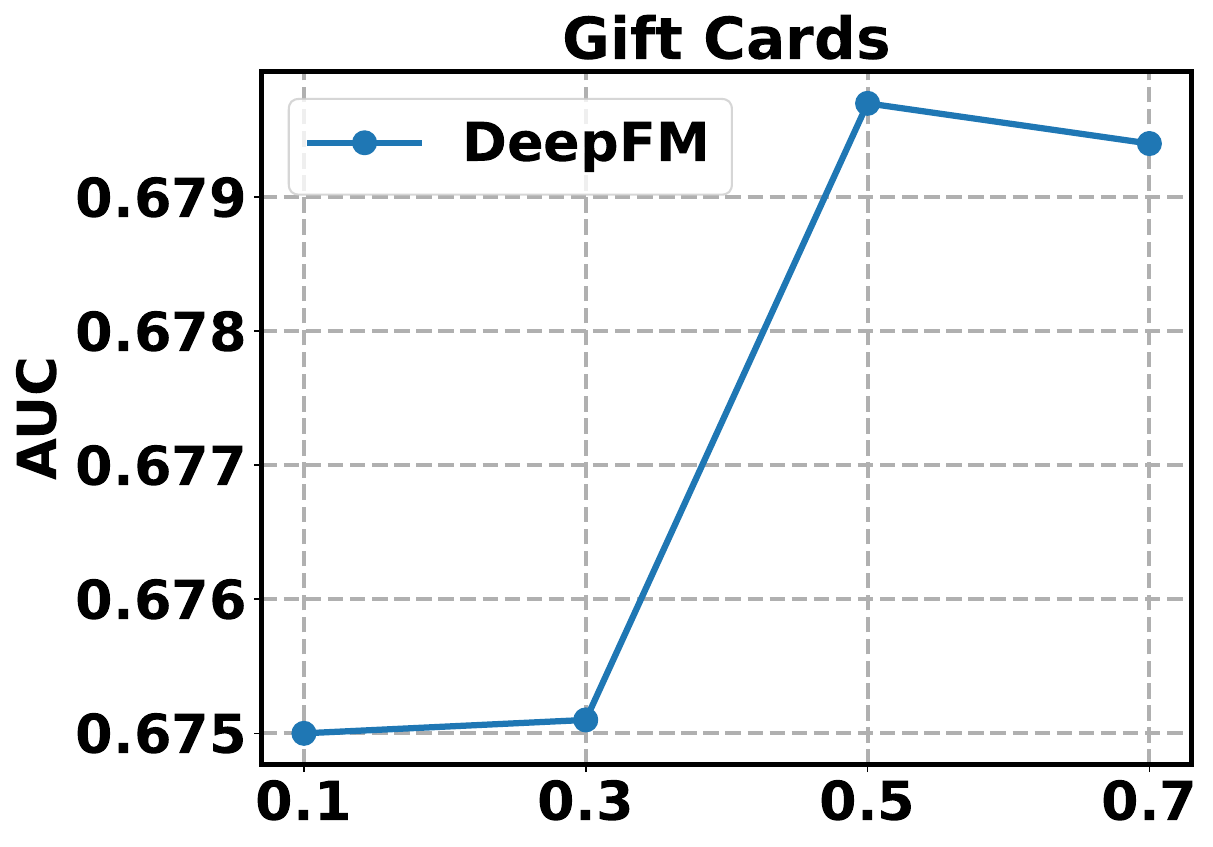}
        \caption{$\quad\quad \lambda_{fm}$}
    \end{subfigure}
    \begin{subfigure}[b]{0.235\textwidth}
    \captionsetup{labelformat=empty}
        \includegraphics[width=\textwidth]{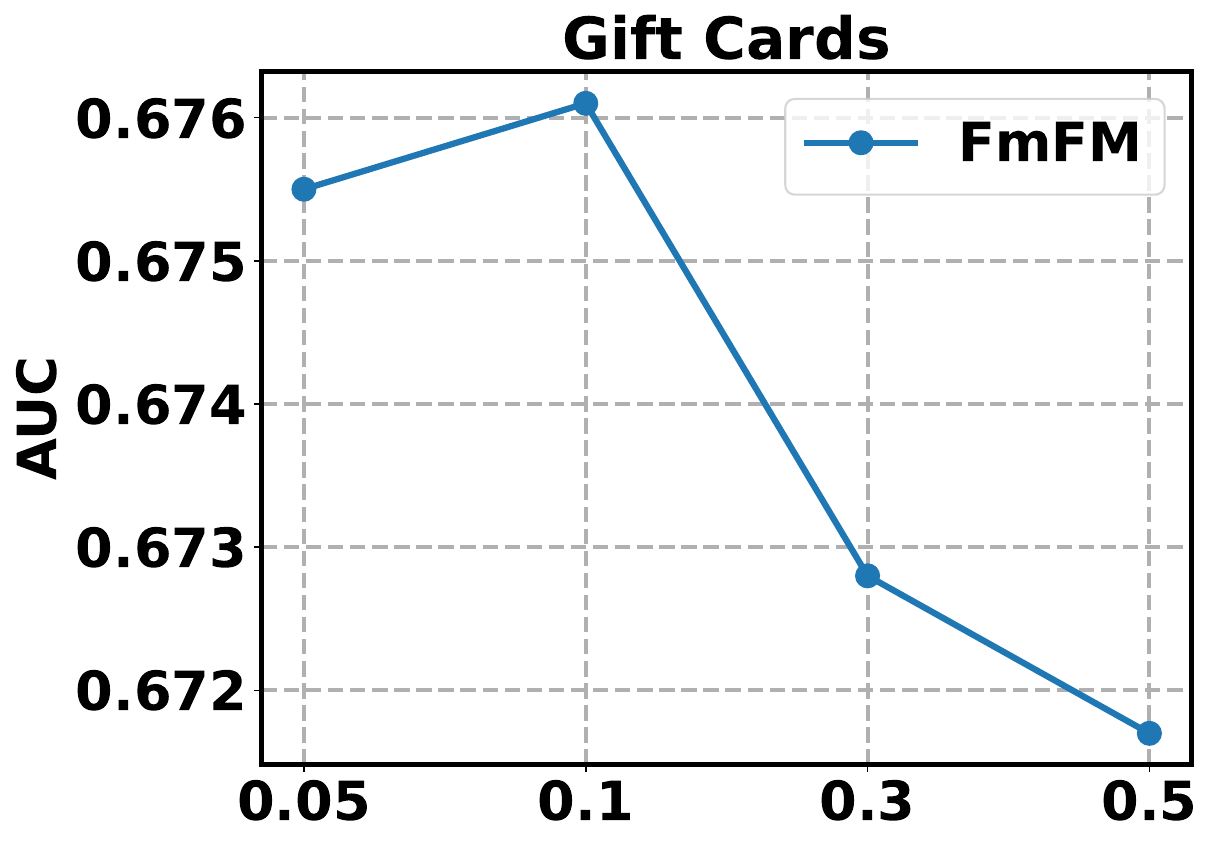}
        \caption{$\quad\quad \lambda_{fm}$}
    \end{subfigure}
    \begin{subfigure}[b]{0.235\textwidth}
    \captionsetup{labelformat=empty}
        \includegraphics[width=\textwidth]{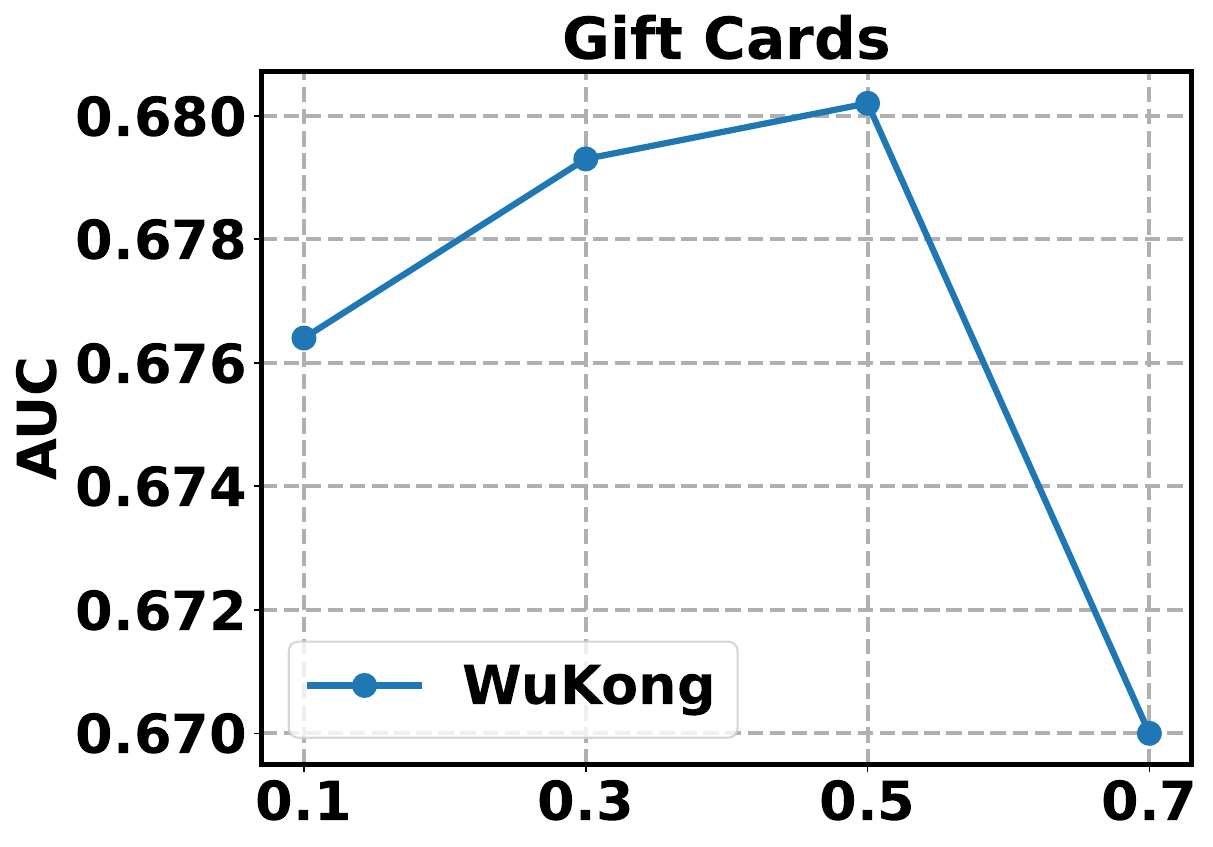}
        \caption{$\quad\quad \lambda_{fm}$}
    \end{subfigure}
    \caption{Sensitivity analysis w.r.t. $\lambda_{kl}$, $\lambda_{fm}$.} 
    \label{fig:Sensitivity}
\end{figure}

Figure~\ref{fig:Sensitivity} illustrates performance of LLaCTR with different hyper-parameters, where $\lambda_{kl}$ and $\lambda_{fm}$ control the effects of Feature Representation Enhancement (FRE) and Feature Interaction Enhancement (FIE) respectively. 
We can observed the general trend is that the model's performance would increase at the beginning and then drop as these parameters increase. This result validates the effectiveness of the Feature Representation Enhancement (FRE) module and Feature Interaction Enhancement (FIE) module. But over-emphasizing the introduced component would incur performance drops as it would relatively decline the contribution from original models. Finely tuning these hyper-parameters for best balance could achieve optimal performance.

\end{document}